\documentclass[a4paper,fleqn,usenatbib]{mnras}

\usepackage[T1]{fontenc}
\usepackage{ae,aecompl}

\usepackage{amssymb}	% Extra maths symbols
\usepackage[font=small,labelfont=bf]{caption}
\usepackage{amssymb,amsmath}
\usepackage{afterpage} 
\usepackage{amsmath} 
\usepackage{booktabs} 
\usepackage{enumerate} 
\usepackage{rotating} 
\usepackage{graphicx} 
\usepackage{subcaption}
\usepackage{hyperref} 
\usepackage{url}
\usepackage{longtable}
\usepackage{makeidx}
\usepackage{url}
\usepackage{natbib}
\usepackage{multirow}
\usepackage{epstopdf}
\usepackage{caption}
\usepackage[section]{placeins}
\usepackage{setspace}
\DeclareCaptionFormat{cont}{#1 (cont.)#2#3\par}
\graphicspath{ {figures/} }

\newcommand{\Sun}{\odot}	% Sun

\title[ULX nebula in NGC 3521]{Discovery and analysis of a ULX nebula in NGC 3521\thanks{Based on observations made with ESO Telescopes at the La Silla Paranal Observatory under programme 0100.D-0660(B).} }

\author[K. M. L\'{o}pez et al.]{
K. M. L\'{o}pez$^{1,2}$\thanks{Contact e-mail: \href{mailto:K.M.Lopez@sron.nl}{K.M.Lopez@sron.nl}},
P. G. Jonker$^{1,2}$,
M. Heida$^{3}$,
M. A. P. Torres$^{4,5,1}$
\newauthor
T. P. Roberts$^{6}$,
D. J. Walton$^{7}$,
D.-S. Moon$^{8}$,
F. A. Harrison$^{3}$
\\
\\
$^{1}$SRON Netherlands Institute for Space Research, 3584 CA Utrecht, The Netherlands
\\
$^{2}$Department of Astrophysics/IMAPP, Radboud University, P.O. Box 9010, 6500 GL Nijmegen, The Netherlands
\\
$^{3}$Space Radiation Laboratory, California Institute of Technology, Pasadena, CA 91125, USA
\\
$^{4}$Instituto de Astrof\'{i}sica de Canarias, E-38200 La Laguna, Tenerife, Spain
\\
$^{5}$Departamento de Astrof\'{i}sica, Universidad de La Laguna, Astrof\'{i}sico Francisco S\'{a}nchez s/n, E-38206 La Laguna, Tenerife, Spain
\\
$^{6}$Centre for Extragalactic Astronomy, Department of Physics, University of Durham, South Road, Durham DH1 3LE, United Kingdom
\\
$^{7}$Institute of Astronomy, Cambridge University, Madingley Road, Cambridge CB3 0HA, United Kingdom
\\
$^{8}$Department of Astronomy and Astrophysics, University of Toronto, Toronto, ON M5S 3H4, Canada
}

\date{Accepted XXX. Received YYY; in original form ZZZ}

\pubyear{2019}

\begin{document}
\label{firstpage}
\pagerange{\pageref{firstpage}--\pageref{lastpage}}
\maketitle

\begin{abstract}
We present Very Large Telescope/X-shooter and {\it Chandra} X-ray observatory/ACIS observations of the ULX
[SST2011] J110545.62+000016.2 in the galaxy NGC 3521. The source identified
as a candidate near-infrared counterpart to the ULX in our previous study
shows an emission line spectrum of numerous recombination and forbidden lines
in the visible and near-infrared spectral regime.
The emission from the candidate counterpart is spatially extended ($\sim$34 pc) and appears
to be connected with an adjacent H{\scshape ii} region, located $\sim$138 pc to the NE. The measured velocities of the
emission lines confirm that both the candidate counterpart and H{\scshape ii} region reside
in NGC 3521. The intensity ratios of the emission lines from the ULX counterpart
show that the line emission originates from the combined effect of shock
and photoionisation of low metallicity (12 + $\log$ (O/H) = 8.19 $\pm$ 0.11) gas.
Unfortunately, there is no identifiable spectral signature directly related to the photosphere of the mass-donor star in our spectrum.
From the archival {\it Chandra} data, we derive the X-ray luminosity of the source in the 0.3--7 keV range to be
$(1.9 \pm 0.8) \times 10^{40}$ erg cm$^{-2}$ s$^{-1}$, almost a factor of four higher than what is previously reported.

\end{abstract}

\begin{keywords}
stars: black holes -- infrared: stars -- X-ray: binaries -- ISM: H{\scshape ii} regions
\end{keywords}

%%%%%%%%%%%%%%%%%%%%%%%%%%%%%%%%%%%%%%%%%%%%%%%%%%

%%%%%%%%%%%%%%%%% BODY OF PAPER %%%%%%%%%%%%%%%%%%

\section{Introduction}
\label{intro}

\begin{table*}
\vspace{5mm}
\begin{center}
\caption{Selected properties of the host galaxy of the observed ULX counterpart.}
\label{tab:galaxies}
\resizebox{\textwidth}{!}{\begin{tabular}{|cccccc|}
\hline
Galaxy & Morphological & Distance & Axis ratio & Inclination & ULX distance to \\
 & type & (Mpc) & & (deg) & galaxy nucleus (kpc)\\
 \hline
NGC 3521 & SAB(rs)bc$^a$ & 14.19 $\pm$ 2.84$^b$  & 0.51$^c$  & 73$^d$ & 10.46 $\pm$ 2.09\\
\hline
\multicolumn{6}{l}{References: $^a$\citet{2011AJ....141...23B}, $^b$\citet{2008AJ....136.2563W}, $^c$\citet{2013AJ....146...86T} and $^d$\citet{2003AJ....125..525J}.}\\
\end{tabular}}
\end{center}
\end{table*}

An ultraluminous X-ray source (ULX) is defined as a point-like, off-nuclear 
source with an X-ray luminosity $L_X$ larger than the Eddington luminosity
for a 10 M$_{\Sun}$ black hole, i.e. $L_X > 10^{39}$ erg s$^{-1}$
\citep{2017ARA&A..55..303K}. To explain these high luminosities,
different possibilities on the nature of the accretors
powering the ULX are considered. The first possibility is the ULX being powered by a stellar mass
compact object with either emitting anisotropically \citep{2001ApJ...552L.109K},
or accreting at super-Eddington rates
\citep{2002ApJ...568L..97B,2003ApJ...586.1280M,2009MNRAS.397.1836G}. Though 
a bona-fide black hole (BH) ULX is yet to be confirmed, several neutron star
ULXs have been discovered in the past years (e.g. 
\citealt{2014Natur.514..202B,2016ApJ...831L..14F,2016arXiv160907375I,2017MNRAS.466L..48I,2018MNRAS.476L..45C}),
in line with the proposed super-Eddington accretion.

The second possibility is that the accretor is a BH more massive than 10M$_{\Sun}$
(e.g. \citealt{2009MNRAS.400..677Z}), i.e. with masses similar 
to the BHs whose merger produced gravitational waves (e.g. \citealt{2016PhRvL.116f1102A}). 
The third possiblity is that the ULX harbors
a BH much more massive than the systems in the first two scenarios, but less massive
than the BHs in the center of most galaxies. These type of BH would accrete at sub-Eddington
rates, have masses between 10$^2$ and 10$^5$M$_{\Sun}$ and are called intermediate mass black holes (IMBHs)
(e.g. \citealt{2011AN....332..392F,2013MNRAS.436.3128M,2016AN....337..448E}).

A reliable way to identify which accretor powers the ULXs is by dynamical
mass measurements. To date, a mass constraint is available for the neutron star ULX M82-X2 
\citep{2014Natur.514..202B}, estimated through the detection of pulsations in the source.
That technique can only be used for ULXs powered by neutron stars, so in the absence of pulsations,
astronomers have focused on detecting the donor star
(e.g. \citealt{2008MNRAS.386..543P,2016MNRAS.459..771H,2018ApJ...854..176V,2019arXiv190401066Q}). 
If detected, spectroscopic observations can be used to constrain the radial velocity amplitude,
and hence, provided the orbital period is determined as well from those data,
the black hole mass function (e.g. \citealt{2011AN....332..367M}). 

On the other hand, other studies have focused on the analysis of the environment in which
a ULX is embedded. Thanks to their high luminosities, ULXs can have a strong effect on their
surroundings, i.e. ionising the gas around them. This is done either by photoionisation due to the
high X-ray and UV luminosity of the ULX, or by shock ionisation caused by jets, or outflows
or disc winds. Examples of ULXs photoionising the nebula surrounding them are Holmberg II X-1
 \citep{2002astro.ph..2488P,2004MNRAS.351L..83K,2011ApJ...731L..32M} and NGC 5408 X-1 
\citep{2006MNRAS.368.1527S}; while ULXs responsible of shock ionising the material around them
are IC 342 X-1 \citep{2006ESASP.604..451G,2012ApJ...749...17C}, Holmberg IX X-1 
\citep{1995ApJ...446L..75M,2006IAUS..230..302G,2011ApJ...731L..32M} 
and M51 ULX-1 \citep{2018MNRAS.475.3561U}.

To distinguish whether a nebula surrounding the ULX is photoionised or shock ionised, one should
look at the spectra. A high $[$S {\scshape ii}$]/H\alpha$ line ratio, i.e. $> 0.4$, is a sign of shock
ionisation \citep{1980A&AS...40...67D}. The electron temperature $T_e$ in a shock
ionised region is higher than in a photoionised region \citep{2018MNRAS.475.3561U}.
The shock ionisation models from \citet{2008ApJS..178...20A} show that a high 
Balmer decrement can be caused by ionising shocks with velocities $> 100$ km s$^{-1}$. Furthermore,
the diagnostic diagrams from \citet{1984ApJ...276..653D} that use the $[$O {\scshape ii}$]\lambda 3727$/H$\beta$,
$[$O {\scshape iii}$]\lambda 5007$/H$\beta$, $[$N {\scshape ii}$]\lambda 6584$/H$\beta$, 
and $[$S {\scshape ii}$]\lambda 6731$/H$\beta$
ratios, could shed some light on the shock velocities of the ionised gas. 
There is, sometimes, the presence of collimated jets (e.g. \citealt{2010Natur.466..209P,2010MNRAS.409..541S,2014Sci...343.1330S}) 
in shock ionised nebulae, evident in the elongated morphology of the nebula with symmetrical lobes.
Nonetheless, in a large number of cases the nebula has both photoionised and shock ionised gas present
(e.g. \citealt{2003MNRAS.342..709R,2007AstBu..62...36A,2007ApJ...668..124A,2018MNRAS.475.3561U}).

We performed a systematic search in the NIR to identify possible counterparts to nearby ULXs \citep{2017MNRAS.469..671L},
where we detected several red supergiant candidates, some of which we followed-up spectroscopically. In this
manuscript, we present the results for one of these sources in particular, [SST2011] J110545.62+000016.2 (hereafter J1105) in 
the galaxy NGC 3521, which turned out to be a nebula. We describe the target in Section~\ref{sample} and the observations 
and data reduction of the spectra in Section~\ref{nirobs}. Our results are presented in Section~\ref{results} and discussed in 
detail in Section~\ref{discussion}. We conclude in Section~\ref{conclusions}.

\section{Target}
\label{sample}

We present the analysis of VLT/X-Shooter spectra of the NIR counterpart to
J1105 in NGC 3521. \citet{2017MNRAS.469..671L} detected the NIR counterpart for 
J1105, with an absolute magnitude $H = -10.93 \pm 0.93$ (consistent with it being a 
red supergiant candidate), assuming a distance to 
NGC 3521 of 14.19 $\pm$ 2.84 Mpc \citep{2008AJ....136.2563W} (see Table~\ref{tab:galaxies}). 
J1105 is located on the northern side of the spiral arms of NGC 3521, 
approximately 10.9 kpc (152\arcsec) from its nucleus.
According to \citet{2018MNRAS.477.1958C}, 
the radial velocity of NGC 3521 at 30\arcsec\ (2.1 kpc) from its center is 175 $\pm$ 50 km s$^{-1}$. 
\citet{2011AJ....141...23B} estimated that at 23.8 kpc (459\arcsec) the radial velocity of 
H{\scshape i} is 220 $\pm$ 10 km s$^{-1}$. It is worth noting that \citet{2011AJ....141...23B} assumed a 
distance to NGC 3521 of 10.7 Mpc. If we assume a distance of 14.19 $\pm$ 2.84 Mpc, 459\arcsec\ 
corresponds to 31.6 kpc. Since we do not have any more information about the radial velocity of
NGC 3521 at other positions, only the two mentioned before, 
we adopt the galactic radial velocity at the location of J1105 as 200 $\pm$ 25 km s$^{-1}$.

\section{Observations and data reduction}
\label{nirobs}

\subsection{X-Shooter data}
\label{xshooter}

We obtained spectra of the counterpart to J1105 with X-Shooter \citep{2011A&A...536A.105V}
on the Very Large Telescope (VLT) UT3 at Cerro Paranal under programme ID 0100.D-0660(B) 
on 2018 March 31. The spectra were taken in service mode.

X-Shooter has three spectroscopic arms that together provide spectral coverage from
the near-UV to the near-IR. We used a slit width of 0.8\arcsec\ in the UVB arm,
0.7\arcsec\ in the VIS arm and 0.6\arcsec\ in the NIR arm, resulting in
resolutions of $\sim$ 6200, $\sim$ 11400 and $\sim$ 8100, respectively. 
Five ABBA nodding sequences with a 5\arcsec\ nodding throw were used. 
The integration times for individual exposures for the UVB,
VIS and NIR arms were 260, 210 and 285 s, respectively. The total exposure times
were 5200 s for the UVB arm, 4200 s for the VIS arm and 5700 s for the NIR arm.
The airmass during our observations was $\sim$ 1.1, and the seeing at the start and end
of our observations was 0.39\arcsec\ and 0.4$\arcsec$, respectively.
The slit was aligned to the parallactic angle, which at the beginning of the observation was
206$^{\circ}$ and at the end of the observation, 180$^{\circ}$. During our observations,
the night was between clear and photometric.

We reduce the data with the {\scshape Reflex} environment
\citep{2013A&A...559A..96F}, processing the observations of J1105
and a standard star observed at similar airmass and close in time to the target.
In the {\scshape Reflex} environment, the workflow designed for X-Shooter data produces
flat-fielded, sky-subtracted, wavelength- and flux- calibrated 1D and 2D spectra. 
The flux calibration is done using the instrument response curve, derived from the
observation of the standard star (see \citealt{2014A&A...568A...9M}).
The rms of the wavelength correction is 0.03 \AA\ for the UVB and VIS regions, and 0.1 \AA\  for the NIR region. 
Since the spectrum for J1105 is an emission line spectrum (see below), we do not perform telluric corrections.
We estimate the slit-losses assuming the slit transmission factor
used by ESO\footnote{\url{https://www.eso.org/observing/etc/doc/formulabook/node18.html}} and a constant seeing,
which yields losses of 8$\%$, 4$\%$ and 2$\%$, in the UVB, VIS and NIR regions, respectively
(see e.g. \citealt{2006SPIE.6269E..2XS}).

\begin{figure}
	    \begin{minipage}{0.5\textwidth}
        \includegraphics[width=\textwidth]{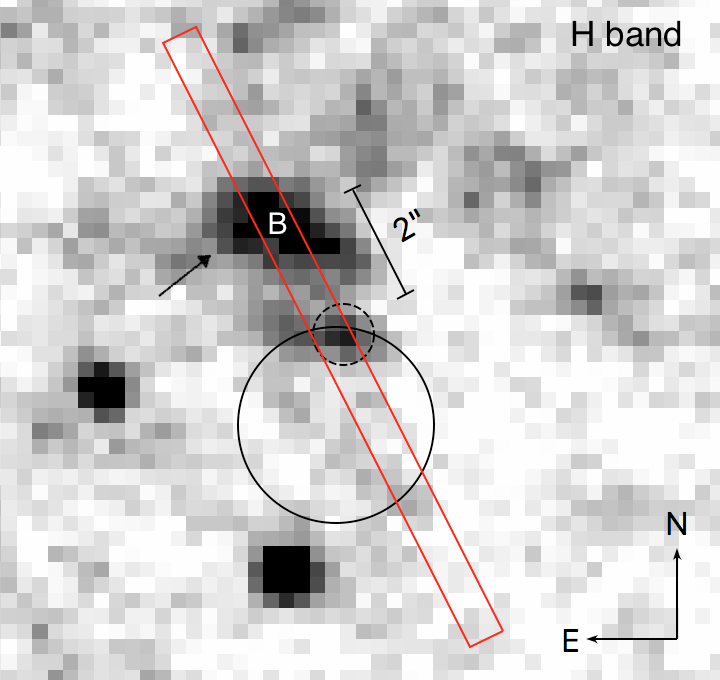}%
        \subcaption{}
        \vspace{0.5cm}     
        \label{fig:slit-image}            
    \end{minipage}   
\caption{LIRIS $H$--band image of J1105 \citep{2017MNRAS.469..671L}. The $0.6\arcsec \times 11\arcsec$ X-Shooter slit is indicated in red, the dashed circle indicates the position of J1105 and the solid circle indicates the position of the ULX to a 3--$\sigma$ confidence level. The extended source B is indicated also with an arrow.}
     \label{fig:j1105}
\end{figure}

To calculate the flux, the full width half maximum (FWHM), velocity dispersion $\sigma$ 
and radial velocity of the emission lines in the three X-Shooter spectral regions, we fit Gaussian curves 
to them. To estimate the uncertainties of these fits we perform a Monte Carlo simulation. We assume
that the noise of the spectrum is Gaussian with a mean equal to zero and a standard deviation equal to 
the value of the rms in each spectral region. We then make 1000 copies of the spectrum of the object,
where each copy is the result of adding a value of this Gaussian distribution of the noise to every data point.
Finally, we fit the emission lines in the resulting spectrum. The standard deviation of each parameter distribution 
is assumed to be its uncertainty. We add statistical uncertainties in quadrature, i.e. the rms of the wavelength 
correction plus the fit uncertainty. 
All values taken from the literature are presented with their published uncertainties.

\subsection{$Chandra$ data}
\label{chandra}

\begin{figure}
        \includegraphics[width=0.4\textwidth, angle=0]{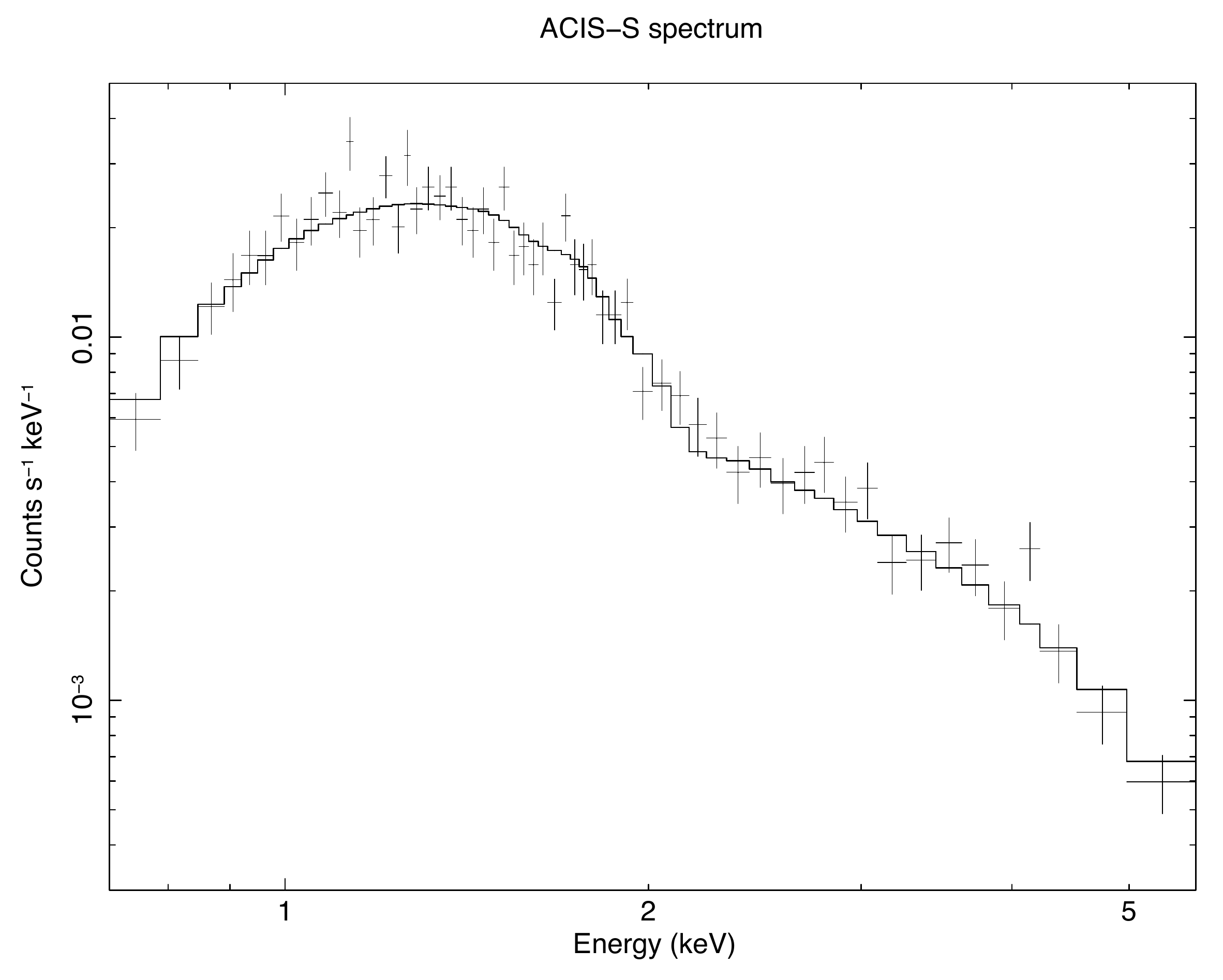}%
	\caption{{\it Chandra} ACIS-S spectrum of J1105 in the 0.3--7 keV energy region fitted with an absorbed power law. For display purposes, the spectrum has been rebinned by maximally 5 bins up to a SNR = 5. The fit was done on the unbinned spectrum. The best--fitting power law index is 2.58 $\pm$ 0.09.}
    \label{fig:xray}  
\end{figure}

We calculate the column of neutral hydrogen $N_H$ and the X-ray flux by analizing
the {\it Chandra}/ACIS archival observation ID 9552 of J1105. 
This observation was performed on 2008 January 28, with an exposure
time of 71.5 ks. We reprocess the event files with the latest
calibration files ({\scshape CALDB} version 4.8.1) using the {\scshape CIAO 4.10} software \citep{2006SPIE.6270E..1VF}.
We extract a source spectrum with the {\scshape ciao} task {\scshape specextract}, using the X-ray position of
J1105 \citep{2011ApJS..192...10L}. We then create a source region of 2\arcsec\ centered on J1105, and a background
region with with inner and outer radii of 10 and 30\arcsec, respectively, both centered on J1105. 
Using the {\sc ciao} task {\it pileup\_map} we derive that pile-up is less than 4\% at the position of the source.
We rebin the extracted source spectrum such that each channel has at least 30 X-ray photons.
We fit the extracted spectra using the {\sc HEASOFT} {\sc XSPEC} tool version 12.10.1 \citep{1996ASPC..101...17A}.
We exclude photons detected outside the range 0.3--7 keV as this energy interval is the best calibrated and 
most sensitive range for {\it Chandra}. We find that a fit-function consisting of an absorbed power law describes the data well 
({\sc phabs $\times$ pegpwr} in {\sc XSPEC}; $\chi^2=51.45$ for 59 degrees of freedom, see Figure~\ref{fig:xray}). 
The best--fitting power law index is 2.58 $\pm$ 0.09 for an N$_H=($0.44 $\pm$ 0.04$)\times 10^{22}$ cm$^{-2}$. 

In order to test potential systematic effects related to the fit-function used to describe the X-ray spectrum, we tried a few other 
well motivated models such as a {\sc phabs $\times$ (diskbb + pegpwr)} to describe the ULX states 
and a {\sc diskpbb} to determine the range of values obtained for N$_H$ given these different continuum models.
Typically, a value of 0.3$\times 10^{22}$ cm$^{-2}$ was obtained using these two fit-functions. 
Therefore, we conclude that the best-fit N$_H$ of 0.4$\times 10^{22}$ cm$^{-2}$ that was obtained from our simple 
absorbed power law model might have been on the high-end of possible values found when more complex fit-functions 
were used to describe the X-ray spectrum, and we add this systematic effect to have a final value of
N$_H = (0.44 \pm 0.14) \times 10^{22}$ cm$^{-2}$. The unabsorbed source 0.3--7 keV flux is then
$($8.3 $\pm$ 1.0$) \times10^{-13}$ erg cm$^{-2}$ s$^{-1}$, which for the distance of NGC 3521, yields 
$L_X = (1.9 \pm 0.8) \times 10^{40}$ erg s$^{-1}$. In contrast, \citet{2011ApJS..192...10L} reported
a flux of 1.92$\times 10^{-13}$ erg cm$^{-2}$ s$^{-1}$, which for the distance of NGC 3521,
translates to $L_X = 4.6 \times 10^{39}$ erg s$^{-1}$, a value almost four times lower than our estimate, 
using a {\sc phabs $\times$ pegpwr} fit with a power law index of 1.7.
It is worth noting that the analysis from \citet{2011ApJS..192...10L} is based on the {\it Chandra} observation
ID 4694, with an exposure time of 9.4 ks, executed on 2004 March 11. Since this observation was performed
four years earlier than the observation we analyse in this manuscript, we can interpret this difference in X-ray luminosities
as variability in the source, which is not uncommon in ULXs (see \citealt{2019MNRAS.486.5709L} and references therein).

\section{Results}
\label{results}

The finding chart for J1105 is shown in Figure~\ref{fig:j1105} where it can be seen that, due to the
orientation of the slit, there are two sources inside it. The NIR counterpart identified in \citet{2017MNRAS.469..671L}
falls near the center of the slit, whereas a brighter extended source is located 2\arcsec (137.6 pc) away (hereafter source B).
We extract the spectra from both these sources and analyse them below. In Figures~\ref{fig:spectra-uvb} 
through~\ref{fig:spectra-nir2} we provide plots of the regions around the most prominent emission lines.

There are several emission lines present in the spectra of both sources (see Table~\ref{tab:lines-wave}).
We measure the radial velocity of the emission line region in the three X-Shooter spectral regions separately 
by fitting Gaussian curves to the lines detected. The average radial velocities found are 238 $\pm$ 10 km s$^{-1}$ 
for J1105 and 235 $\pm$ 4 km s$^{-1}$ for source B, which are both consistent with the radial velocity of NGC 3521 of 
200 $\pm$ 25 km s$^{-1}$, confirming that both objects are not foreground objects.

We calculate the intrinsic FWHM (FWHM$_{\rm int}$) of the emission lines. In order to do this, we estimate the instrumental FWHM
(FWHM$_{{\rm ins}}$), i.e. the width that a delta function line would have due to the instrument, given by the resolution of X-Shooter 
in the respective arms. As indicated in Section ~\ref{xshooter}, the resolution R in the UVB, VIS and NIR arms is 6200, 
11400 and 8100, respectively, and it is defined as R = $\frac{\lambda}{\Delta \lambda}$, where $\Delta \lambda$ is equivalent to
FWHM$_{{\rm ins}}$. Note that this value is wavelength dependent.
We then calculate the FWHM$_{\rm int}$ via the equation 
FWHM$_{\rm int} = ($FWHM$_{\rm obs}^2 - $ FWHM$_{\rm ins}^2)^{1/2}$, where FWHM$_{\rm obs}$ 
is the observed FWHM, i.e. the FWHM we derive by fitting Gaussians to the emission lines. We present the FWHM$_{\rm int}$
in Table~\ref{tab:lines-wave}.

\subsection{Nebular source B}
\label{sourceB}

\begin{figure*}
        \includegraphics[width=\textwidth]{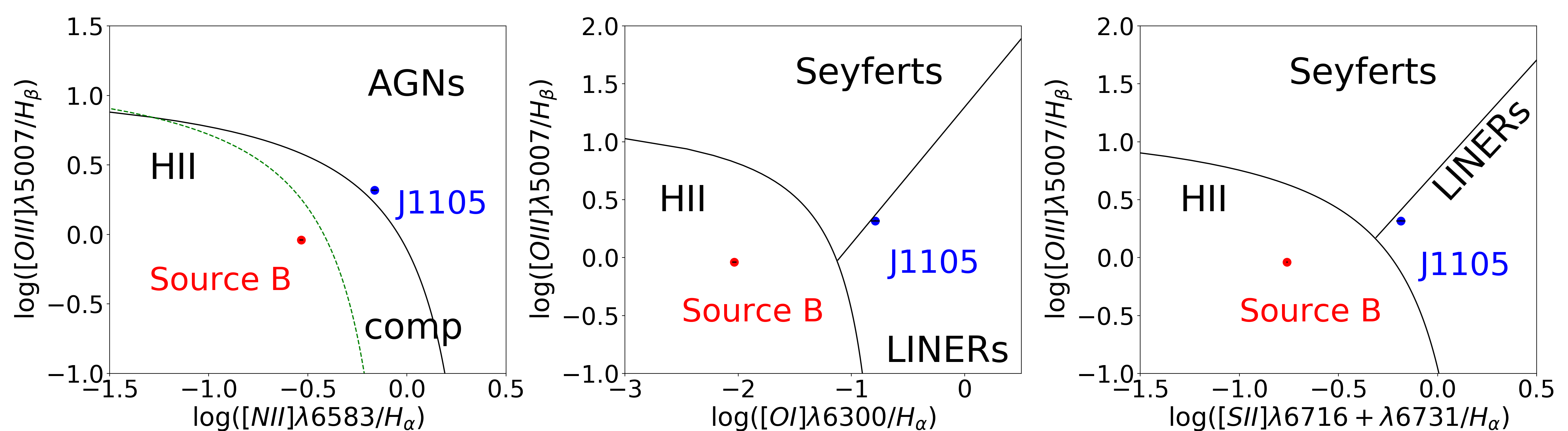}%
	\caption{Baldwin-Phillips-Terlevich diagrams for J1105. The green dashed and the black solid lines in the left panel mark the empirical and the theoretical maximum line ratios for H{\scshape ii} regions from \citet{2003MNRAS.346.1055K} and \citet{2006MNRAS.372..961K}, respectively. The red circle represents the line flux ratios at the location of source B, while the blue circle represents the line flux ratios at the location of J1105. The size of error bars is of the order of the size of the symbols. The difference between the dereddened and uncorrected values for the line flux ratios is of the order of the size of the symbols.}
    \label{fig:bpt}  
\end{figure*} 

\subsubsection{UVB spectra}
\label{uvb}

Faint continuum emission is detected in the UVB and VIS regions only of the spectrum from source B.
We find the $[$O {\scshape ii}$] \lambda \lambda$3727, 3729 doublet, 
H$\beta$, $[$Ne {\scshape iii}$]\lambda$3868 and $[$O {\scshape iii}$]\lambda \lambda$4956, 5007 lines. 
We also detect the $[$Fe {\scshape i}$]\lambda$3744 line, although the Gaussian fit to this feature did not 
converge at the position of source B. So, we deem this a marginal detection for source B.

We also find the $[$He {\scshape i}$]\lambda$3888, H$\delta$, H$\epsilon$ and H$\gamma$ lines. 
The He {\scshape ii} $\lambda$4686 emission line, observed in several ULX nebulae (e.g.
\citealt{2002astro.ph..2488P,2004MNRAS.351L..83K,2014ApJ...797L...7G}) is not detected with a 2--$\sigma$
upper limit of 2.26$\times 10^{-19}$ erg cm$^{-2}$ s$^{-1}$.

\subsubsection{VIS spectra}
\label{vis}

We detect $[$O {\scshape i}$]\lambda$6300, $[$N {\scshape ii}$]\lambda$6548, 
H$\alpha$, $[$N {\scshape ii}$] \lambda$6583, the $[$S {\scshape ii}$] \lambda \lambda$6716, 6731 doublet,
$[$S {\scshape iii}$]\lambda \lambda$9069, 9530 and Pa$\delta$ emission. 
Additionally, we detect emission from $[$He {\scshape i}$]\lambda$6678, $[$Ar {\scshape iii}$]\lambda \lambda$7135, 7751, 
$[$O {\scshape ii}$]\lambda \lambda \lambda$7320, 7329, 7330, Pa10, Pa9, Pa$\epsilon$ and the temperature 
sensitive auroral $[$S {\scshape iii}$]\lambda$6312. The latter allows us to calculate the 
electron temperature $T_e$ and density $n_e$, and ion abundances.

Additionally, we detected an emission line only in the spectrum from source B, at the wavelength 7325 \AA\ , which
we were not able to identify.

\subsubsection{NIR spectra}
\label{nir}

We detect He {\scshape i}, $^3$He {\scshape i}, Pa$\beta$, Pa$\alpha$, Pa$\gamma$, 
$[$Fe {\scshape ii}$]\lambda$12566, $[$Fe {\scshape ii}$]\lambda$16435, $[$N {\scshape iii}$]$, 
Br19, Br16, Br13, Br12, Br11, Br10 and Br9 emission.

\subsubsection{Emission line analysis}
\label{analysis}

\begin{figure}
        \includegraphics[width=0.5\textwidth]{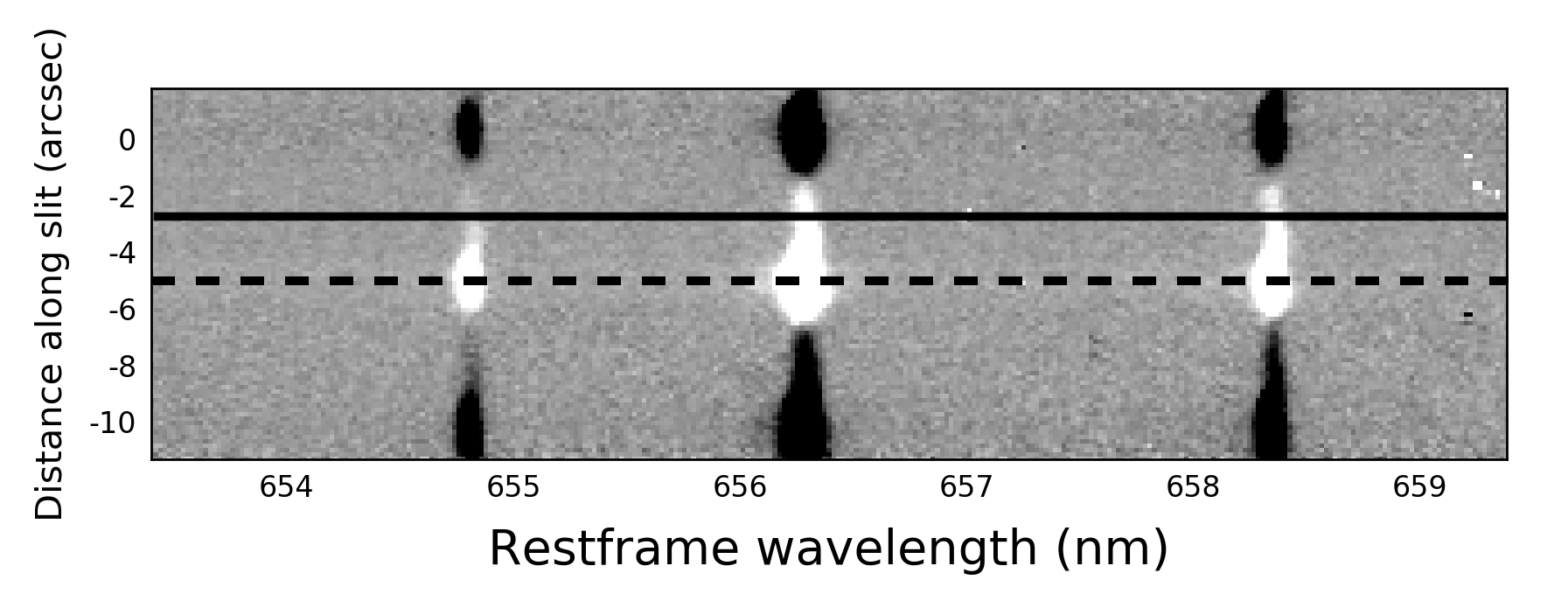}%
		\caption{2D image of the VIS part of the X-Shooter spectrum of the NIR counterpart to J1105. Here it can be seen how the $[$N {\scshape ii}$]\lambda$6548, H$\alpha$ and $[$N {\scshape ii}$] \lambda$6583 emission lines are extended at the position of source B (indicated by the dashed line) along both the spatial and wavelength/velocity axis. The black solid horizontal line indicates the position of the NIR counterpart J1105 on the detector.}
		\label{fig:2dspectra}
\end{figure}

Using the results from the Gaussian fits to the emission lines, we calculate the line flux ratios
$\log( [$O {\scshape iii}$]/$H$\beta )$, $\log( [$N {\scshape ii}$]/$H$\alpha )$, $\log( [$S {\scshape ii}$]/$H$\alpha )$
and $\log( [$O {\scshape i}$]/$H$\alpha )$. As can be seen in Figure~\ref{fig:bpt}, the flux line rations
for source B are consistent with those of an H{\scshape ii} region \citep{2003MNRAS.346.1055K,2006MNRAS.372..961K}.
Moreover, in Figure ~\ref{fig:2dspectra} we show that the emission lines associated with source B are extended.

The $n_e$ for source B was calculated using both the $[$O {\scshape ii}$]$ and $[$S {\scshape ii}$]$ line ratios 
\citep{1949ApJ...109...42A,2002A&A...382..282C,2006agna.book.....O} using the {\scshape TEMDEN}
task from the {\scshape ANALYSIS} package in {\scshape Iraf}. 
The calculations give $n_e[$O {\scshape ii}$]$ = 50 $\pm$ 15 cm$^{-3}$ and
$n_e[$S {\scshape ii}$]$ = 62 $\pm$ 24 cm$^{-3}$, consistent with each other. These $n_e$ values suggest
that the H{\scshape ii} region associated with source B is a low-density region \citep{2006agna.book.....O}.
With these values and the 
$[$S {\scshape iii}$]\lambda$9069, 9530/$[$S {\scshape iii}$]\lambda$6312 intensity ratio 
(e.g. \citealt{2006agna.book.....O}) we calculated a value of $T_e[$S {\scshape iii}$]$ = 8130 $\pm$ 1213 K. 
This method is applicable to densities up to 10$^5$ cm$^{-3}$ (i.e. negligible collisional deexcitations) 
and assuming an isothermal environment (see \citealt{2006agna.book.....O}).

We were also able to calculate the $[$O {\scshape ii}$]$, $[$O {\scshape iii}$]$, $[$S {\scshape ii}$]$ and 
$[$S {\scshape iii}$]$ ion abundances, following the equations from \citet{1992MNRAS.255..325P},
adequate for low to moderate $n_e$ (i.e. up to 10$^3$ - 10$^4$ cm$^{-3}$, \citealt{2006agna.book.....O}). 
Additionally, we determined the empirical oxygen and sulfur abundance indicators, $R_{23}$ and $S_{23}$, respectively 
\citep{1979MNRAS.189...95P,1996MNRAS.280..720V}, the ionisation parameter u \citep{1991MNRAS.253..245D}
and the radiation softness parameter $\eta '$ \citep{1988MNRAS.231..257V}. The latter is an indicator on the temperature
of the ionizing stars in an H{\scshape ii} region. The derived values are given in Table~\ref{tab:abundance}.

\subsection{The ULX counterpart}
\label{ulxcp}

\subsubsection{UVB spectra}
\label{uvbcp}

We detect the $[$O {\scshape ii}$] \lambda \lambda$3727, 3729 doublet, $[$Fe {\scshape i}$]\lambda$3744,
H$\beta$, $[$Ne {\scshape iii}$]\lambda$3868 and $[$O {\scshape iii}$]\lambda \lambda$4956, 5007 lines. 
Addtionally, we detect emission lines from $[$Fe {\scshape ii}$]\lambda \lambda$4436, 4452.
For this source we also do not detect He {\scshape ii} $\lambda$4686 emission, with a 2--$\sigma$
upper limit of 2.12$\times 10^{-19}$ erg cm$^{-2}$ s$^{-1}$.

\subsubsection{VIS spectra}
\label{viscp}

We detect $[$O {\scshape i}$]\lambda$6300, $[$N {\scshape ii}$]\lambda$6548, 
H$\alpha$, $[$N {\scshape ii}$] \lambda$6583, the $[$S {\scshape ii}$] \lambda \lambda$6716, 6731 doublet,
$[$S {\scshape iii}$]\lambda \lambda$9069, 9530 and $[$Fe {\scshape i}$]\lambda$10048 emission. 
The auroral $[$S {\scshape iii}$]\lambda$6312 line was not detected at the position of J1105 with a 2--$\sigma$ upper limit 
of 2.58$\times 10^{-18}$ erg cm$^{-2}$ s$^{-1}$. This implies that we can only place limits on the $T_e$ and $n_e$
parameters.

\subsubsection{NIR spectra}
\label{nircp}
 
The lines present in the NIR data for J1105 are He {\scshape i}, $^3$He {\scshape i}, Pa$\beta$, 
$[$Fe {\scshape ii}$]\lambda$16435 and Pa$\alpha$. Faint continuum emission is detected only for J1105. 
However, the spectrum is too faint to allow us determine if the continuum is stellar.

To verify whether the flux detected with X-Shooter is consistent with the apparent 
magnitudes given by \citet{2017MNRAS.469..671L} in the $H$--band, we estimate the 
apparent magnitudes from the flux calibrated spectra using the equation:

\begin{equation}
m_H = -2.5\log\bigg(\frac{F_H}{F_{H,0}}\bigg)
\end{equation}

where $m_H$ is the apparent magnitude in the $H$-band, $F_H$ is the flux of the emission lines
detected in the wavelength range $1500 - 1750$ nm (corresponding to the $H$-band) and 
$F_{H,0}$ is the reference flux ($1.883 \pm 0.038\times 10^{-6}$ erg cm$^{-2}$ s$^{-1}$, \citealt{2003AJ....126.1090C}).
We assume $R_V = 3.1$ \citep{1999PASP..111...63F} and $R_{H} = 0.46 \pm 0.01$ \citep{2013MNRAS.430.2188Y}
to correct the observed flux for extinction. 

The total integrated flux of J1105 in the $H$--band is 1.78$\times 10^{-15}$ erg cm$^{-2}$ s$^{-1}$, 
which is equivalent to an apparent magnitude of 22.54 $\pm$ 2.01, consistent within 2-$\sigma$ with our reported value 
of $H$ = 19.83 $\pm$ 0.06 mag \citep{2017MNRAS.469..671L}.

\subsubsection{Emission line analysis}
\label{analysis}

We also calculate the line flux ratios $\log( [$O {\scshape iii}$]/$H$\beta )$, $\log( [$N {\scshape ii}$]/$H$\alpha )$, 
$\log( [$S {\scshape ii}$]/$H$\alpha )$ and $\log( [$O {\scshape i}$]/$H$\alpha )$ for J1105 and, as can be seen in 
Figure~\ref{fig:bpt}, they are consistent with those of Low-ionisation nuclear emission-line regions (LINERs) and 
Active Galactic Nuclei (AGNs) \citep{2003MNRAS.346.1055K,2006MNRAS.372..961K}.

We were not able to calculate ion abundance for J1105, as they are dependent on $T_e$-, $n_e$- and 
$[$S {\scshape iii}$]\lambda$6312, and the latter was not detected for J1105.
We can, however, obtain a lower limit on the temperature at the location of J1105, assuming that it is a photoionised
region, like source B. We follow the same procedure as for source B but we found that the values for
the $[$O {\scshape ii}$]$ and $[$S {\scshape ii}$]$ line ratios are at the limit of $n_e \rightarrow 0$. In fact,
the $[$S {\scshape ii}$]$ line ratio is outside the range allowed by photoionisation models. We derive a value for
$n_e[$O {\scshape ii}$]$ of 5 $\pm$ 5 cm$^{-3}$, that, combined with the upper limit of the
$[$S {\scshape iii}$]\lambda$9069, 9530/$[$S {\scshape iii}$]\lambda$6312 intensity ratio, yields
$T_e[$S {\scshape iii}$] > $ 5270 K.

\section{Discussion}
\label{discussion}

\subsection{The ULX counterpart}
\label{j1105ulxdisc}

\citet{2017MNRAS.469..671L} identified a possible counterpart to J1105 as a potential red supergiant based on its photometry.
However, using VLT/X-Shooter spectroscopy we find that it has an emission line spectrum, whose line ratios 
place it in the LINER region of the Baldwin-Phillips-Terlevich (BPT) diagram. 
These line ratios are different from those of a nearby region called source B (see Figure~\ref{fig:bpt}), and
could be interpreted as due to the X-ray emission ionising the part of the nebular source B that surrounds the ULX. 
The effects of the bright ULX on its environment has been seen in other H{\scshape ii} regions surrounding (or close to) a ULX (e.g.
\citealt{2005ApJ...633L.101M,2015MNRAS.453.3510H}).
Since the presence of the ULX seems to affect the line flux ratios of J1105, and its radial velocity is 
consistent with the radial velocity of NGC 3521 at that position, we deem it likely that the ULX and the nebular
source are physically related. Therefore, we refer to J1105 as the counterpart of the ULX although the mass
donor star has not been detected so far.

\begin{table*}
\vspace{5mm}
\hspace{5mm}
\begin{center}
\caption{Theoretical and restframe wavelengths, integrated flux and intrinsic FWHM of the emission lines in the UVB, VIS and NIR data for both source B and J1105.}
\label{tab:lines-wave}
\resizebox{\textwidth}{!}{\setlength{\tabcolsep}{3pt}
\begin{tabular}{|cccccccc|}
\hline
& Theoretical & \multicolumn{2}{c}{Restframe} &  \multicolumn{2}{c}{Integrated flux} &  \multicolumn{2}{c}{FWHM$_{\rm int}$}\\
Ion & wavelength & \multicolumn{2}{c}{wavelength (\AA)$\dagger$} &  \multicolumn{2}{c}{($\times 10^{-17}$ erg cm$^{-2}$ s$^{-1}$)} & \multicolumn{2}{c}{(\AA)}\\
 & (\AA) * & Source B & J1105 & Source B & J1105 & Source B & J1105 \\
\hline
$[$O \scshape {ii}$]$ & 3726.032 $\pm$ 0.010 & 3726.21 $\pm$ 0.04 & 3726.11 $\pm$ 0.03 & (1.63 $\pm$ 0.01)$\times 10^{2}$ & 11.3 $\pm$ 0.1 & 0.59 $\pm$ 0.03 & 0.84 $\pm$ 0.03 \\
$[$O \scshape {ii}$]$ & 3728.815 $\pm$ 0.010 & 3728.88 $\pm$ 0.03 & 3728.91 $\pm$ 0.03 & (2.29 $\pm$ 0.03)$\times 10^{2}$ & 16.9 $\pm$ 0.1 & 0.63 $\pm$ 0.03 & 0.77 $\pm$ 0.03 \\
$[$Fe {\scshape i}$]$ & 3744.10236 $\pm$ 0.0002 & - & 3744.22 $\pm$ 0.03 & - & 7.81 $\pm$ 0.72 & - & 0.65 $\pm$ 0.03 \\
$[$Ne \scshape {iii}$]$ & 3868.76 $\pm$ 0.10 & 3868.79 $\pm$ 0.01 & 3868.8 $\pm$ 0.1 & 4.83 $\pm$ 0.01 & 3.1 $\pm$ 0.1& 0.7 $\pm$ 0.1 & 0.7 $\pm$ 0.1 \\
$[$He \scshape {i}$]$ & 3888.41607 $\pm$ 0.00021 & 3888.54 $\pm$ 0.04 & - & 16.2 $\pm$ 0.1 & - & 0.86 $\pm$ 0.03 & - \\
H$\epsilon$ & 3970.0788 $\pm$ 0.0022 & 3970.14 $\pm$ 0.03 & - & 19.4 $\pm$ 0.1 & - & 0.68 $\pm$ 0.03 & - \\
H$\delta$ & 4101.7415 $\pm$ 0.0024 & 4101.81 $\pm$ 0.03 & - & 39.9 $\pm$ 0.1 & - & 0.72 $\pm$ 0.03 & - \\
H$\gamma$ & 4340.471 $\pm$ 0.003 & 4340.49 $\pm$ 0.03 & 4340.24 $\pm$ 0.03 & 77.6 $\pm$ 0.3 & 2.59 $\pm$ 1.09 & 0.77 $\pm$ 0.03 & 1.04 $\pm$ 0.03 \\
$[$Fe {\scshape ii}$]$ & 4436.9204 $\pm$ 0.0003 & - & 4436.7 $\pm$ 0.1 & - & 0.55 $\pm$ 0.04 & - & 0.7 $\pm$ 0.1 \\
$[$Fe {\scshape ii}$]$ & 4452.09585 $\pm$ 0.00012 & - & 4452.1 $\pm$ 0.1 & - & 1.23 $\pm$ 0.01 & - & 0.7 $\pm$ 0.1 \\
H$\beta$ & 4861.333 $\pm$ 0.003 & 4861.39 $\pm$ 0.03 & 4861.54 $\pm$ 0.04 & (1.95 $\pm$ 0.01)$\times 10^{2}$ & 8.86 $\pm$ 0.08 & 0.87 $\pm$ 0.03 & 0.87 $\pm$ 0.03 \\
$[$O \scshape {iii}$]$ & 4958.911 $\pm$ 0.010 & 4959.01 $\pm$ 0.03 & 4958.88 $\pm$ 0.03 & 60.0 $\pm$ 0.2 & 6.71 $\pm$ 0.19 & 0.83 $\pm$ 0.03 & 1.12 $\pm$ 0.03 \\
$[$O \scshape {iii}$]$ & 5006.843 $\pm$ 0.010 & 5006.81 $\pm$ 0.03 & 5006.88 $\pm$ 0.03 & (1.78 $\pm$ 0.02)$\times 10^{2}$ & 18.4 $\pm$ 0.4 & 0.83 $\pm$ 0.03 & 0.83 $\pm$ 0.03 \\
\hline
$[$O \scshape {i}$]$ & 6300.304 $\pm$ 0.010 & 6300.5 $\pm$ 0.2 & 6300.46 $\pm$ 0.01 & 7.11 $\pm$ 0.40 & 5.83 $\pm$ 0.03 & 0.8 $\pm$ 0.5 & 0.79 $\pm$ 0.04 \\
$[$S \scshape {iii}$]$ & 6311.21 $\pm$ 0.10 & 6311.6 $\pm$ 0.1 & - & 1.86 $\pm$ 0.64 & - & 1.5 $\pm$ 0.5 & - \\
$[$N \scshape {ii}$]$ & 6548.05 $\pm$ 0.10 & 6548.11 $\pm$ 0.01 & 6547.87 $\pm$ 0.01 & 74.6 $\pm$ 1.0 & 7.03 $\pm$ 0.07 & 0.76 $\pm$ 0.03 & 0.91 $\pm$ 0.03 \\
H$\alpha$ & 6562.819 $\pm$ 0.007 & 6562.85 $\pm$ 0.01 & 6562.75 $\pm$ 0.03 & (7.57 $\pm$ 0.07)$\times 10^{2}$ & 36.0 $\pm$ 0.1 & 0.88 $\pm$ 0.03 & 0.96 $\pm$ 0.03 \\
$[$N \scshape {ii}$]$ & 6583.45 $\pm$ 0.10 & 6583.48 $\pm$ 0.01 & 6583.38 $\pm$ 0.03 & (2.22 $\pm$ 0.05)$\times 10^{2}$ & 24.7 $\pm$ 0.1 & 0.85 $\pm$ 0.03 & 0.86 $\pm$ 0.03 \\
$[$S \scshape {ii}$]$ & 6716.44 $\pm$ 0.01 & 6716.62 $\pm$ 0.01 & 6716.63 $\pm$ 0.01 & 76.0 $\pm$ 0.7 & 15.0 $\pm$ 0.3 & 0.79 $\pm$ 0.04 & 0.98 $\pm$ 0.01 \\
$[$S \scshape {ii}$]$ & 6730.816 $\pm$ 0.010 & 6730.92 $\pm$ 0.01 & 6730.83 $\pm$ 0.01 & 55.7 $\pm$ 0.5 & 8.51 $\pm$ 0.04 & 0.81 $\pm$ 0.04 & 0.97 $\pm$ 0.04 \\
He \scshape {i} & 6678.1517 $\pm$ 0.0003 & 6678.11 $\pm$ 0.03 & - & 7.4 $\pm$ 0.1 & - & 0.81 $\pm$ 0.03 & - \\
$[$Ar \scshape {iii}$]$ & 7135.79 $\pm$ 0.04 & 7135.84 $\pm$ 0.03 & - & 18.6 $\pm$ 0.1 & - & 0.79 $\pm$ 0.03 & - \\
$[$O \scshape {ii}$]$ & 7319.99 & 7320.11 $\pm$ 0.01 & - & 4.2 $\pm$ 0.1 & - & 0.94 $\pm$ 0.03 & - \\
? & - & 7325.71 $\pm$ 0.03 & - & 4.46 $\pm$ 0.03 & - & 0.64 $\pm$ 0.03 & - \\
$[$O \scshape {ii}$]$ & 7329.67 $\pm$ 0.1 & 7329.59 $\pm$ 0.01 & - & 2.25 $\pm$ 0.02 & - & 0.65 $\pm$ 0.03 & - \\
$[$O \scshape {ii}$]$ & 7330.73 $\pm$ 0.1 & 7330.84 $\pm$ 0.03 & - & 2.28 $\pm$ 0.01 & - & 0.83 $\pm$ 0.03 & - \\
$[$Ar \scshape {iii}$]$ & 7751.11 $\pm$ 0.05 & 7751.14 $\pm$ 0.03 & - & 4.31 $\pm$ 0.01 & - & 0.82 $\pm$ 0.03 & - \\
Pa10 & 9014.910 $\pm$ 0.007 & 9015.03 $\pm$ 0.03 & - & 6.69 $\pm$ 0.65 & - & 1.15 $\pm$ 0.04 & - \\
$[$S \scshape {iii}$]$ & 9069.0 $\pm$ 1 & 9068.84 $\pm$ 0.02 & 9069.74 $\pm$ 0.03 & 46.2 $\pm$ 1.3 & 2.38 $\pm$ 0.22 & 1.05 $\pm$ 0.01 & 2.1 $\pm$ 0.1 \\
Pa9 & 9229.014 $\pm$ 0.007 & 9229.04 $\pm$ 0.01 & - & 7.99 $\pm$ 0.96 & - & 1.18 $\pm$ 0.01 & - \\
$[$S \scshape {iii}$]$ & 9530.6 $\pm$ 1 & 9530.81 $\pm$ 0.02 & 9530.64 $\pm$ 0.01 & (1.21 $\pm$ 0.25)$\times 10^{2}$ & 8.21 $\pm$ 0.19 & 1.22 $\pm$ 0.03 & 1.69 $\pm$ 0.04 \\
Pa$\epsilon$ & 9545.969 $\pm$ 0.006 & 9545.47 $\pm$ 0.06 & - & 13.7 $\pm$ 2.0 & - & 1.3 $\pm$ 0.1 & - \\
$[$Fe \scshape {i}$]$& 10048.5845 $\pm$ 0.0014 & - & 10048.0 $\pm$ 0.1 & - & 3.9 $\pm$ 2.5 & - & 0.7 $\pm$ 0.3 \\
Pa$\delta$ & 10049.369 $\pm$ 0.007 & 10049.9 $\pm$ 0.1 & - & 19 $\pm$ 2 & - & 1.6 $\pm$ 0.1 & - \\
\hline
He \scshape {i} & 10830.33977 $\pm$ 0.00005 & 10830.4 $\pm$ 0.1 & 10830.4 $\pm$ 0.1 & 70.3 $\pm$ 0.3 & 3.65 $\pm$ 0.30 & 1.9 $\pm$ 0.1 & 2.1 $\pm$ 0.1 \\
Pa$\gamma$ & 10938.086 $\pm$ 0.01 & 10938.2 $\pm$ 0.1 & - & 34.7 $\pm$ 0.1 & - & 1.7 $\pm$ 0.1 & - \\
$[$Fe \scshape {ii}$]$ & 12566.7688 $\pm$ 0.0009 & 12567.3 $\pm$ 0.4 & - & 8.28 $\pm$ 1.94 & - & 2.5 $\pm$ 0.6 & - \\
$^3$He \scshape {i} & 12785.50005 $\pm$ 0.00006 & 12785.9 $\pm$ 0.5 & 12785.1 $\pm$ 0.1 & 2.85 $\pm$ 0.01 & 0.75 $\pm$ 0.07 & 1.6 $\pm$ 1.2 & 1.8 $\pm$ 0.1 \\
$[$N \scshape {iii}$]$ & 12809 $\pm$ 1 & 12809.5 $\pm$ 0.6 & - & 3.063 $\pm$ 0.001 & - & 1.5 $\pm$ 1.6 & - \\
Pa$\beta$ & 12818.07 $\pm$ 0.01 & 12818.2 $\pm$ 0.1 & 12818.1 $\pm$ 0.1 & 70.6 $\pm$ 0.1 & 2.22 $\pm$ 0.01 & 1.8 $\pm$ 0.1 & 2.1 $\pm$ 0.1 \\
Br19 & 15260.555 $\pm$ 0.009 & 15259.6 $\pm$ 0.1 & - & 1.12 $\pm$ 0.05 & - & 3.5 $\pm$ 0.1 & - \\
Br16 & 15556.466 $\pm$ 0.015 & 15556.0 $\pm$ 0.1 & - & 1.37 $\pm$ 0.07 & - & 3.1 $\pm$ 0.1 & - \\
Br13 & 16109.330 $\pm$ 0.013 & 16109.3 $\pm$ 0.1 & - & 2.39 $\pm$ 0.09 & - & 3.4 $\pm$ 0.1 & - \\
Br12 & 16407.210 $\pm$ 0.016 & 16406.3 $\pm$ 0.1 & - & 2.07 $\pm$ 0.21& - & 3.1 $\pm$ 0.2 & - \\
$[$Fe \scshape {ii}$]^{\ddagger}$ & 16435.5271 $\pm$ 0.0019 & 16434.3 $\pm$ 0.1 & 16436.3 $\pm$ 0.1 & 5.70 $\pm$ 0.87 & 0.92 $\pm$ 0.10 & 5.7 $\pm$ 0.1 & 2.1 $\pm$ 0.1 \\
Br11 & 16806.538 $\pm$ 0.017 & 16807.0 $\pm$ 0.1 & - & 4.09 $\pm$ 0.09 & - & 2.3 $\pm$ 0.1 & - \\
Br10 & 17362.127 $\pm$ 0.018 & 17363.1 $\pm$ 0.1 & - & 4.47 $\pm$ 0.16 & - & 2.9 $\pm$ 0.1 & - \\
Br9 & 18174.141 $\pm$ 0.020 & 18173.7 $\pm$ 0.1 & - & 5.35 $\pm$ 1.64 & - & 2.9 $\pm$ 0.1 & - \\
Pa$\alpha$ & 18750.976 $\pm$ 0.025 & 18751.4 $\pm$ 0.1 & 18751.1 $\pm$ 0.1 & 91.8 $\pm$ 3.1 & 3.99 $\pm$ 0.01 & 2.7 $\pm$ 0.1 & 2.6 $\pm$ 0.1 \\
\hline
\multicolumn{8}{l}{*Wavelengths are indicated in air, not vacuum, and the reference values are taken from \citet{NIST_ASD}.}\\ 
\multicolumn{8}{l}{$^{\dagger}$Corrected for heliocentric motion and peculiar and radial velocity of NGC 3521.}\\ 
\multicolumn{8}{l}{$^{\ddagger}$The properties of this line are possibly contaminated by the presence of a sky emission line close to the observed wavelength.}\\ 
\end{tabular}}
\end{center}
\end{table*}

We calculate the extinction at the position of J1105 using the N$_H$ we derived by analising the
{\it Chandra} archival observations, with observation ID 9552, of the ULX. 
We obtained N$_H = (0.44 \pm 0.14) \times 10^{22}$ atoms cm$^{-2}$, which for a typical value of $R_V = 3.1$, 
translates to $E(B-V) = 0.76 \pm 0.24$. If we deredden the H$\alpha$ and H$\beta$ line fluxes with this extinction 
and then calculate the Balmer decrement, we get unphysical values, i.e. H${\alpha}/$H${\beta} < 2$
(values only seen in environments with $n_e > 10^9$ cm$^{-3}$, \citealt{1974ApJ...192..199A}).
This can be explained if the X-ray emission from the ULX and the emission from the ionised gas comes from different regions,
implying that there is more neutral gas in the line of sight towards the X-ray source than towards the gas
responsible for the H$\alpha$ and H$\beta$ emission lines.

In light of this, we decide to calculate the extinction using the Balmer decrement for J1105 (Table~\ref{tab:lineratios}).
Assuming a standard Milky Way gas-to-dust ratio ($R_V = 3.1$, \citealt{Fitzpatrick_1999}) and an intrinsic
Balmer decrement of H${\alpha}/$H${\beta} = 3.05$ (valid for a typical optically thick nebula in the low density limit, 
with $T_e \sim$ 5,000 K), we derive $E(B-V) = 0.27 \pm 0.03$. The N$_H$ corresponding to this extinction is 
N$_H = (0.15 \pm 0.02) \times 10^{22}$ atoms cm$^{-2}$.
We then deredden the key fluxes with this two values of extinction and 
report them in Table~\ref{tab:lineratios}, where it can be seen that they still place J1105 in the LINER region
of the BPT diagram.

\begin{table}
\vspace{5mm}
\begin{center}
\caption{Electron temperature and density, and ion abundances for source B and J1105}
\label{tab:abundance}
\begin{tabular}{|ccc|}
\hline
Parameter & Source B & J1105 \\
 \hline
T$_e$ & 8130 $\pm$ 1213 K & $>$ 5270 K \\
$n_e[$O {\scshape ii}$]$ & 50 $\pm$ 15 cm$^{-3}$ & 5 $\pm$ 5 cm$^{-3}$\\
$n_e[$S {\scshape ii}$]$ & 62 $\pm$ 24 cm$^{-3}$ & -\\
12 + $\log$ (O {\scshape ii}/H {\scshape ii}) & 7.93 $\pm$ 0.14 & -\\
12 + $\log$ (O {\scshape iii}/H {\scshape ii}) & 7.83 $\pm$ 0.19 & -\\
12 + $\log$ (O/H) & 8.19 $\pm$ 0.11 & -\\
12 + $\log$ (S {\scshape ii}/H {\scshape ii}) & 6.43 $\pm$ 0.14 & -\\
12 + $\log$ (S {\scshape iii}/H {\scshape ii}) & 6.69 $\pm$ 0.15 & -\\
12 + $\log$ (S/H) & 6.97 $\pm$ 0.12 & -\\
$\log$ R$_{23}$ & 0.52 $\pm$ 0.03 & -\\
$\log$ S$_{23}$ & 0.182 $\pm$ 0.001& -\\
$\log U$ & $-$2.84 $\pm$ 0.08 & -\\
$\log \eta '$ & 0.30 $\pm$ 0.01 & -\\
\hline
\end{tabular}
\end{center}
\end{table}

From the abundances derived for the extended source B nearby (Table~\ref{tab:abundance}), we conclude that
J1105 resides in a low-metallicity environment. This is in line with previous findings that ULXs are
preferentially found in low-metallicity environments (e.g.
\citealt{2011ApJ...741...10K,2011MNRAS.416.1844W,2013ApJ...769...92P,2016ApJ...818..140B}).
As stated above, the dereddened line flux ratios of J1105 place it in the LINER region of the BPT diagram.
Though the power source behind the emission of a LINER is under debate, several explanations have
been considered, e.g. photoionisation by hot young stars \citep{1985MNRAS.213..841T,1992ApJ...399L..27S},
low-luminosity active galactic nuclei \citep{1983ApJ...269L..37H,1983ApJ...264..105F},
planetary nebulae \citep{2000AJ....120.1265T} or exposed cores of evolved stars \citep{2013A&A...558A..43S,2016MNRAS.461.3111B},
or shock-ionization \citep{1980A&A....87..152H,1996ASPC..103...44D}
We further investigate whether the latter is the case for J1105.
We first analyse the $[$S {\scshape ii}$]$/H$\alpha$ ratio, which distinguishes between shock ionised and 
photoionised regions: shock ionised regions have $[$S {\scshape ii}$]$/H$\alpha > $ 0.4 \citep{1980A&AS...40...67D}.
The dereddened ratio for source B is 0.17, a typical value for H{\scshape ii} regions 
\citep{1995AJ....110..739L,2008MNRAS.383.1175P},
whereas J1105 has a dereddened ratio of 0.64. Moreover, in the line flux ratio diagrams often used to distinguish between 
supernova remnants
with shocks and H{\scshape ii} regions by \citet{1995AJ....110..739L}, J1105 falls in the supernova remnants region 
while source B is consistent with being an H{\scshape ii} region (Table~\ref{tab:lineratios}).
Another clue on whether J1105 is shock ionised could come from the $[$O {\scshape ii}$]\lambda 3727$/H$\beta$, 
$[$N {\scshape ii}$]\lambda 6584$/H$\beta$, and $[$S {\scshape ii}$]\lambda 6731$/H$\beta$ ratios. 
Our dereddened values (Table~\ref{tab:lineratios}) are consistent with
shock velocities between 50 and 80 km s$^{-1}$ \citep{1984ApJ...276..653D}. When we compare with 
the FWHMs from J1105 (see Table~\ref{tab:lines-wave}), we find that the average FWHM for this source is 50 $\pm$ 1 km s$^{-1}$,
whereas for source B is 45 $\pm$ 1 km s$^{-1}$. Hence, our results are consistent with the diagrams from \citet{1984ApJ...276..653D}.

We estimate the mechanical power $P_{{\rm jet}}$ of J1105. According to \citet{1977ApJ...218..377W}
and \citet{2012ApJ...749...17C}, for a nebula which is shock ionised,
$P_{{\rm jet}} \sim 2.85 L_{{\rm rad}}$, where $P_{{\rm jet}}$ is the power that inflates the 
nebula bubble and $L_{{\rm rad}}$ is the radiative luminosity of the bubble.
Following \citet{1996ApJS..102..161D,2007ApJ...668..124A,2008ApJS..178...20A} and \citet{2012ApJ...749...17C}
we know that $L_{{\rm H}\beta} = 6.53 \times 10^{-3} v_{100}^{-0.59}L_{{\rm rad}}$ where $L_{{\rm H}\beta}$ is the luminosity 
from H$\beta$ emission and $v_{100}$ is the shock velocity in units of 100 km s$^{-1}$. This means
that $P_{{\rm jet}} = 437 v_{100}^{0.59}L_{{\rm H}\beta}$.
To calculate $L_{{\rm H}\beta}$, we deredden our $H\beta$ flux using $E(B-V) = 0.27 \pm 0.03$, which at the distance of NGC 3521, 
is equivalent to $L_{{\rm H}\beta} = 5.25 \times 10^{36}$ erg s$^{-1}$. 
Hence, for shock velocities between 50 and 80 km s$^{-1}$, $P_{{\rm jet}}$ is $(1.52 - 2.01) \times 10^{39}$, i.e. 
almost one-tenth of our derived X-ray luminosity of the ULX and 0.3--0.4 times the X-ray
luminosity derived by \citet{2011ApJS..192...10L}. In comparison, for ULX IC 342-X1, $P_{{\rm jet}} \sim \frac{1}{20}L_X$ \citep{2012ApJ...749...17C},
for M51 ULX-1, $P_{{\rm jet}} \sim L_X$ \citep{2018MNRAS.475.3561U} and for S26, $P_{{\rm jet}} \sim 10^{4}L_X$ \citep{2010Natur.466..209P}.

As we have seen in other nebulae surrounding ULXs
(e.g. \citealt{2005A&A...431..847L,2006IAUS..230..278F,2007AstBu..62...36A,2007ApJ...668..124A}),
both photoionised and shock ionised gas is present. We consider the possibility that 
shocks are not the only ionisation source and the X-ray emission from J1105 is photoionising the region as well. 
To investigate this scenario further, we calculate the Str\"{o}mgren radius $R_S$ \citep{1939ApJ....89..526S},
i.e. the radius of the photoionised region, if any, expected to be created by J1105 (see e.g. \citealt{2003MNRAS.342..709R}). 
First we need to calculate the hydrogen ionising flux $Q$, which correlates with $L_{{\rm H}\beta}$ as 
$L_{{\rm H}\beta} = 4.8 \times 10^{-13} Q$ erg $s^{-1}$ \citep{2006agna.book.....O}.
This gives a value of $Q = 1.1 \times 10^{49}$ photons s$^{-1}$, and assuming a filling factor $\epsilon$ between 0.01 and 0.5, 
a typical range for most H{\scshape ii} regions \citep{2006agna.book.....O} and $T_e = 10000$ K, 
we get 30 $< R_S <$ 111 pc. In comparison, assuming spherical symmetry, we estimate a radius of $\sim 34$ pc for 
J1105 from its appearance in our $H$-band image.
So, for a $\epsilon < 0.35$, it would be possible for J1105 to be photoionising the entirety of the region. However,
we do not need the entirety of the region to be photoionised, since we are inferring that the total ionisation is a combination
of X-ray photoionisation and shock ionisation. Alternatively, a wind blown off the accretion disc
of the ULX can also cause shock ionisation \citep{2007AstBu..62...36A,2017ARA&A..55..303K}.
In fact, powerful disc winds have been observed in a few ULXs 
(e.g. \citealt{2016ApJ...826L..26W,2016Natur.533...64P,2017AN....338..234P,2018MNRAS.479.3978K}).

\subsection{Nebular source B}
\label{j1105disc}

The nebular ion abundances for source B are consistent with those seen in other H{\scshape ii} regions
(see e.g. \citealt{2003RMxAC..17..205P,2005MNRAS.361.1063P}).
The $n_e$ value is at the low density end, whereas typical $n_e$ in H{\scshape ii} regions are of the order
of 10$^2$ cm$^{-3}$ \citep{2006agna.book.....O}. The density that we obtain, within uncertainties, is seen
in nebulae like NGC 281 and NGC 7000 \citep{1981A&A....94..238R,2006agna.book.....O}.

\citet{Moustakas_2010} studied the metallicity of NGC 3521 in
an area enclosing $\sim$ 49$\%$ of the integrated light, up to 3.9 kpc from the nucleus.
The oxygen abundance they calculated is 12 + $\log$(O/H) = 9.08 $\pm$ 0.06, 
which could be consistent with our value of 8.19 $\pm$ 0.11, since metallicity in spiral galaxies decreases
with distance to the nucleus \citep{1981ARA&A..19...77P,2000MNRAS.311..329D,2016A&A...588A..91M},
and source B is located 10.5 kpc from the nucleus. This is further confirmed with the $\log R_{23}$ parameter,
i.e. the oxygen vs. H$\beta$ abundance indicator. \citet{Moustakas_2010} get $\log R_{23} = 0.32 \pm 0.20$ 
and we get $\log R_{23} = 0.52 \pm 0.03$, indicative of a decrease in the oxygen abundance 
(see \citealt{2005MNRAS.361.1063P}). The value we calculate for $\log S_{23}$ (the sulfur vs. H$\beta$ abundance indicator) 
is also consistent with our value for 12 + $\log$ (O/H).
\citet{2018MNRAS.477.1958C} measured the metallicity of NGC 3521 between its nucleus and 30\arcsec\ 
from the nucleus. At the latter position, they get a value of [Z/H] = $-$0.3. In comparison, we calculate 
values of [O/H] = $-$0.65 and [S/H] = $-$0.31\footnote{Using $\log O_{\Sun} = 8.83$ and $\log S_{\Sun} = 7.27$ 
(e.g. \citealt{2003Obs...123..320B}).}, consistent with a metallicity decrease.

\citet{Moustakas_2010} also calculated the ionisation parameter for NGC 3521 (in an area enclosing 49$\%$ of the integrated light),
getting a value of $\log U =-2.89 \pm 0.25$, consistent with our measurement, since $\log U$ was found to maintain
a practically constant value with distance to the nucleus in some galaxies (i.e. M33, \citealt{1988MNRAS.235..633V}).
Additionally, they found a Balmer decrement of
H${\alpha}/$H${\beta} = 5.78 \pm 0.50$, whereas we found a value of 3.88, indicative of lower
extinction \citep{1983ApJ...267..119M,1991MNRAS.253..245D,2005MNRAS.361.1063P} at the position of source 
B than at 3.9 kpc from the nucleus and consistent with the extinction decreasing as a function of distance 
to the nucleus of a galaxy \citep{2016ApJ...819..152C}.

With the Balmer decrement we could estimate the extinction at the position
of source B, assuming a standard Milky Way gas-to-dust ratio ($R_V = 3.1$, \citealt{Fitzpatrick_1999}). 
We assume an intrinsic H${\alpha}/$H${\beta} = 2.87$, valid for a typical optically thick nebula
(recombination case B, low density limit, and $T_e \sim$ 10,000 K \citealt{2006agna.book.....O})
and we obtain $E(B-V) = 0.28 \pm 0.01$. We deredden the emission line fluxes from source B and
calculate again the line ratios, finding that they are still consistent with source B being an H {\sc ii} region
(see Table~\ref{tab:lineratios}).

The $T_e$ we find for the H{\scshape ii} region source B is also within the range of observed temperatures in 
H{\scshape ii} regions, i.e. 7000--10000 K
(e.g. \citealt{1992ApJ...389..305O,1997ApJ...489...63G,1998MNRAS.295..401E,2003ApJ...591..801K}).
Lastly, our value of the radiation softness parameter $\eta '$ along with 12 + $\log$ (O/H) correspond to an H{\scshape ii} region
ionised by a young hot stellar population, with an effective temperature $T_{eff} = 40,000 K$ 
(for $\log U = -2.6$, \citealt{1999ApJ...510..104B,2000ApJ...531..813V}). This is in line with previous findings that ULXs often reside
close to young OB associations (e.g. \citealt{2008A&A...486..151G,2011ApJ...734...23G,2011MNRAS.418L.124V,2012ApJ...758...28J}).

\begin{table*}
\vspace{5mm}
\begin{center}
\caption{Observed and dereddened emission line flux ratios and Balmer decrements for source B and J1105. For the dereddening we used $E(B-V) = 0.28 \pm 0.01$ and $E(B-V) = 0.27 \pm 0.03$ for source B and J1105, respectively (see text).}
\label{tab:lineratios}
%\resizebox{\textwidth}{!}{
\begin{tabular}{|cccccc|}
\hline
Line & \multicolumn{2}{c}{Source B} & \multicolumn{2}{c}{J1105}\\
flux ratio & Observed & Dereddened & Observed & Dereddened\\
 \hline
$\log( [$O {\scshape iii}$]\lambda 5007/$H$\beta )^a$ & $-$0.04 $\pm$ 0.02 & $-$0.06 $\pm$ 0.02 & 0.32 $\pm$ 0.01 & 0.30 $\pm$ 0.01\\
$\log( [$O {\scshape i}$]\lambda 6300/$H$\alpha )^a$& $-$2.03 $\pm$ 0.02 & $-$2.02 $\pm$ 0.02 & $-$0.79 $\pm$ 0.01 & $-$0.78 $\pm$ 0.0\\
$\log( [$N {\scshape ii}$]\lambda 6583/$H$\alpha )^a$& $-$0.53 $\pm$ 0.02 & $-$0.54 $\pm$ 0.02 & $-$0.16 $\pm$ 0.01 & $-$0.17 $\pm$ 0.01\\
$\log( [$S {\scshape ii}$]\lambda 6716 + \lambda 6731/$H$\alpha )^{a,b,c}$& $-$0.76 $\pm$ 0.02 & $-$0.77 $\pm$ 0.02 & $-$0.185 $\pm$ 0.004 & $-$0.194 $\pm$ 0.004\\ 
$\log( [$N {\scshape ii}$]\lambda 6548+ \lambda 6583$/H$\alpha )^c$ & $-$0.41 $\pm$ 0.02 & $-$0.41 $\pm$ 0.02 & $-$0.055 $\pm$ 0.003 & $-$0.056 $\pm$ 0.003\\
$\log( [$S {\scshape iii}$]\lambda 9069 + \lambda 9530$/H$\alpha )^c$ & $-$0.7 $\pm$ 0.2 & $-$0.8 $\pm$ 0.2 & $-$0.53 $\pm$ 0.03 & $-$0.65 $\pm$ 0.03\\
$\log( [$O {\scshape ii}$]\lambda 3727$/H$\beta )^d$ & N/A & N/A & 0.28 $\pm$ 0.01 & 0.39 $\pm$ 0.01\\
$\log( [$N {\scshape ii}$]\lambda 6583$/H$\beta )^d$ & N/A & N/A & 0.45 $\pm$ 0.01 & 0.31 $\pm$ 0.01\\
$\log( [$S {\scshape ii}$]\lambda 6731$/H$\beta )^d$ & N/A & N/A & $-$0.02 $\pm$ 0.01 & $-$0.16 $\pm$ 0.01\\
\hline
H${\alpha}/$H${\beta}$ & 3.88 $\pm$ 0.01 & 2.81 $\pm$ 0.01& 4.06 $\pm$ 0.12 & 2.97 $\pm$ 0.12\\
H${\gamma}/$H${\beta}$ & 0.398 $\pm$ 0.003 & 0.452 $\pm$ 0.003 & 0.29 $\pm$ 0.03 & 0.33 $\pm$ 0.03\\
H${\delta}/$H${\beta}$ & 0.205 $\pm$ 0.001 & 0.246 $\pm$ 0.00& - & -\\
H${\epsilon}/$H${\beta}$ & 0.10 $\pm$ 0.01 & 0.12 $\pm$ 0.01 & - & -\\
\hline
\multicolumn{6}{l}{{\bf Notes: }$^a$Diagnostic line flux ratios in the BPT diagram, to distinguish AGNs, LINERs and H {\sc ii} regions} \\
\multicolumn{6}{l}{\citep{2003MNRAS.346.1055K,2006MNRAS.372..961K}. $^b$Line ratio that distinguishes between shock and photoionised}\\
\multicolumn{6}{l}{regions \citep{1980A&AS...40...67D}. $^c$Diagnostic line flux ratios to distinguish supernova remnants with shocks}\\
\multicolumn{6}{l}{from H {\sc ii} regions \citep{1995AJ....110..739L}. $^d$Line flux ratios indicative of shock velocities, from the diagrams}\\
\multicolumn{6}{l}{\citet{1984ApJ...276..653D}. N/A = not applicable.}\\
\end{tabular}%}
\end{center}
\end{table*}

\section{Conclusions}
\label{conclusions}

We took X-Shooter spectra of a candidate NIR counterpart to the ULX J1105 in NGC 3521. 
We obtained spectra of both the counterpart of the ULX and a nearby HII region. We detect
several emission lines from these two spectra. The line flux ratios
from the extended source B classify it as an H{\scshape ii} region, while the line flux ratios from J1105 place it in the
LINER region of the BPT diagram. Source B has the characteristics of a typical low density H{\scshape ii} region, with a
$T_e = 8130 \pm 1213$ K and $n_e = 50 \pm 15$ cm$^{-3}$. It is a low metallicity region, i.e. 12 + $\log$ (O/H) = 8.19 $\pm$ 0.11.
This is consistent with the fact that several ULXs are found primarily in low metallicity environments and
close to young OB associations.
J1105 is a different object. The fact that the line ratios at the position of this source are different than source B
is interpreted as being caused by the ULX. So we establish that the ULX is located in NGC 3521 although
we do not detect the mass donor star. We calculate $L_X$ from the $Chandra$ observation of the
ULX and derive a value significantly higher than the one reported before by \citet{2011ApJS..192...10L},
implying source variability.
Additionally, we derive a lower limit of $T_e \sim 5270$ K for J1105 and its line ratios indicate that this source could be shock ionised. 
The diagnostic diagrams from \citet{1984ApJ...276..653D}
suggest that the shocks have low velocity, i.e. between 50 and 80 km s$^{-1}$. This is further confirmed by the average
FWHM from J1105 (50 $\pm$ 1 km s$^{-1}$).
However, the luminosity of the H$\beta$ line seems to indicate that the ULX is photoionising the nebula.
Hence, we argue that the nebula is not only shock ionised but also photoionised by the X-ray radiation from the ULX.
The shock ionisation could be driven by either a disc wind or a jet.

\section*{Acknowledgements}
We would like to thank the anonymous referee whose valuable comments helped improve the quality of this manuscript.
KML would like to thank Davide Lena for valuable discussions.
PGJ and KML acknowledge funding from the European Research Council under ERC Consolidator Grant agreement no 647208. TPR acknowledges funding from STFC as part of the consolidated grant ST/L00075X/1. MAPT acknowledges support via a Ram\'on y Cajal Fellowship (RYC-2015-17854). MAPT also acknowledges support by the Spanish Ministry of Economy, Industry and Competitiveness under grant AYA2017-83216-P.
DJW acknowledges support from an STFC Ernest Rutherford Fellowship. KML 
We have made use of the SIMBAD database, operated at CDS, Strasbourg, France; of the NASA/IPAC Extragalactic Database (NED) which is operated by the Jet Propulsion Laboratory, California Institute of Technology, under contract with the National Aeronautics and Space Administration; and of data obtained from the Chandra Data Archive and the Chandra Source Catalog, and software provided by the Chandra X-ray Center (CXC) in the application packages CIAO, ChIPS, and Sherpa.

%%%%%%%%%%%%%%%%%%%%%%%%%%%%%%%%%%%%%%%%%%%%%%%%%%



\bibliographystyle{mnras}
\bibliography{library}

%%%%%%%%%%%%%%%%% APPENDICES %%%%%%%%%%%%%%%%%%%%%

\appendix

\section{Spectra}

\begin{figure*}
        \includegraphics[width=\textwidth]{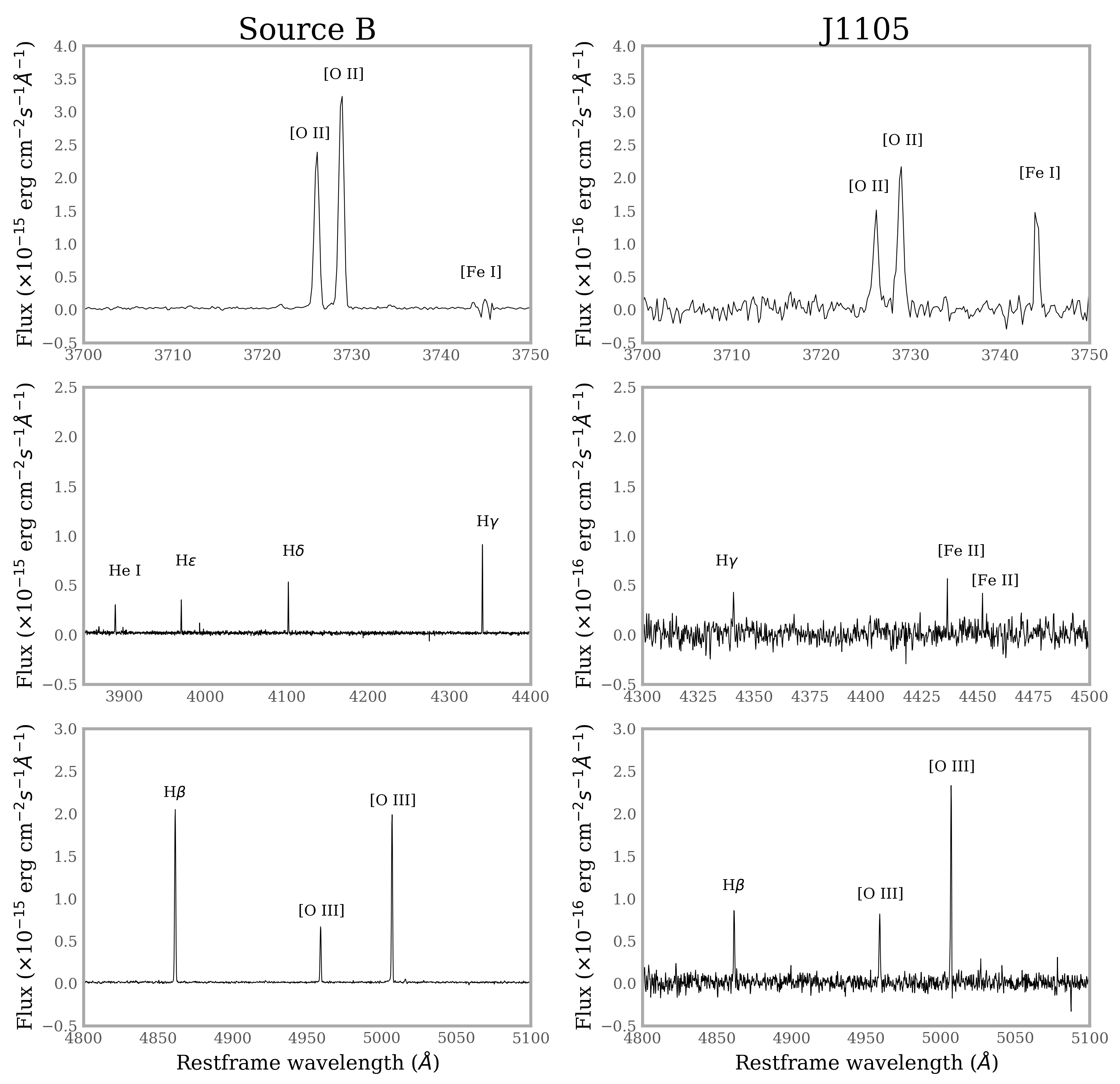}%
		\caption{Selected spectra showing strong emission lines from the UVB range of the X-Shooter spectra from Source B (left) and J1105 (right). The line classification is shown in the plot.}
        \label{fig:spectra-uvb}
\end{figure*}
        
\begin{figure*}
        \includegraphics[width=\textwidth]{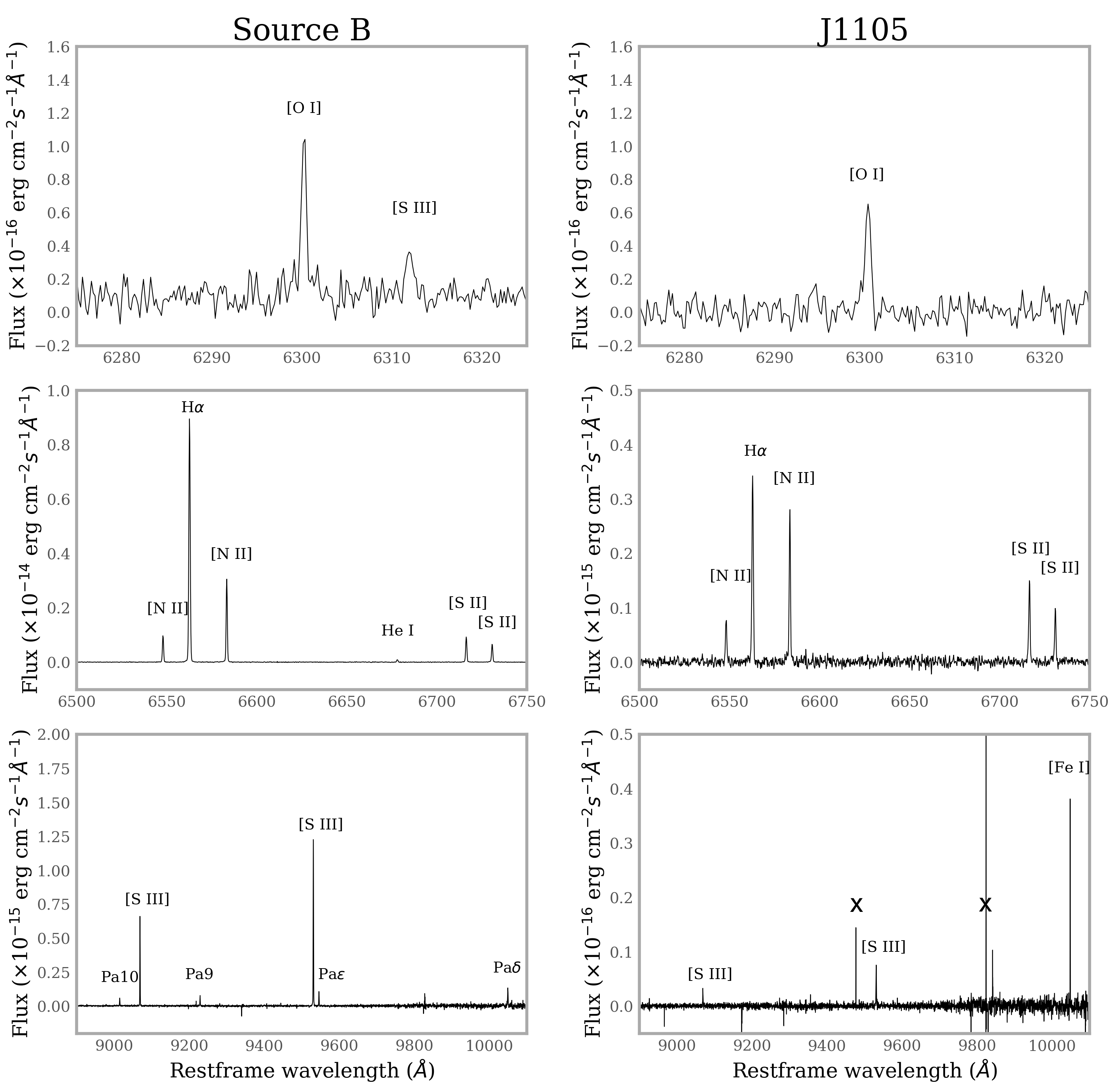}%
		\caption{Selected spectra showing strong emission lines from the VIS range of the X-Shooter spectra from Source B (left) and J1105 (right). The line classification is shown in the plot. Residuals of sky lines subtraction or spurious lines are marked with an x.}
        \label{fig:spectra-vis}
\end{figure*}

\begin{figure*}
        \includegraphics[width=0.5\textwidth]{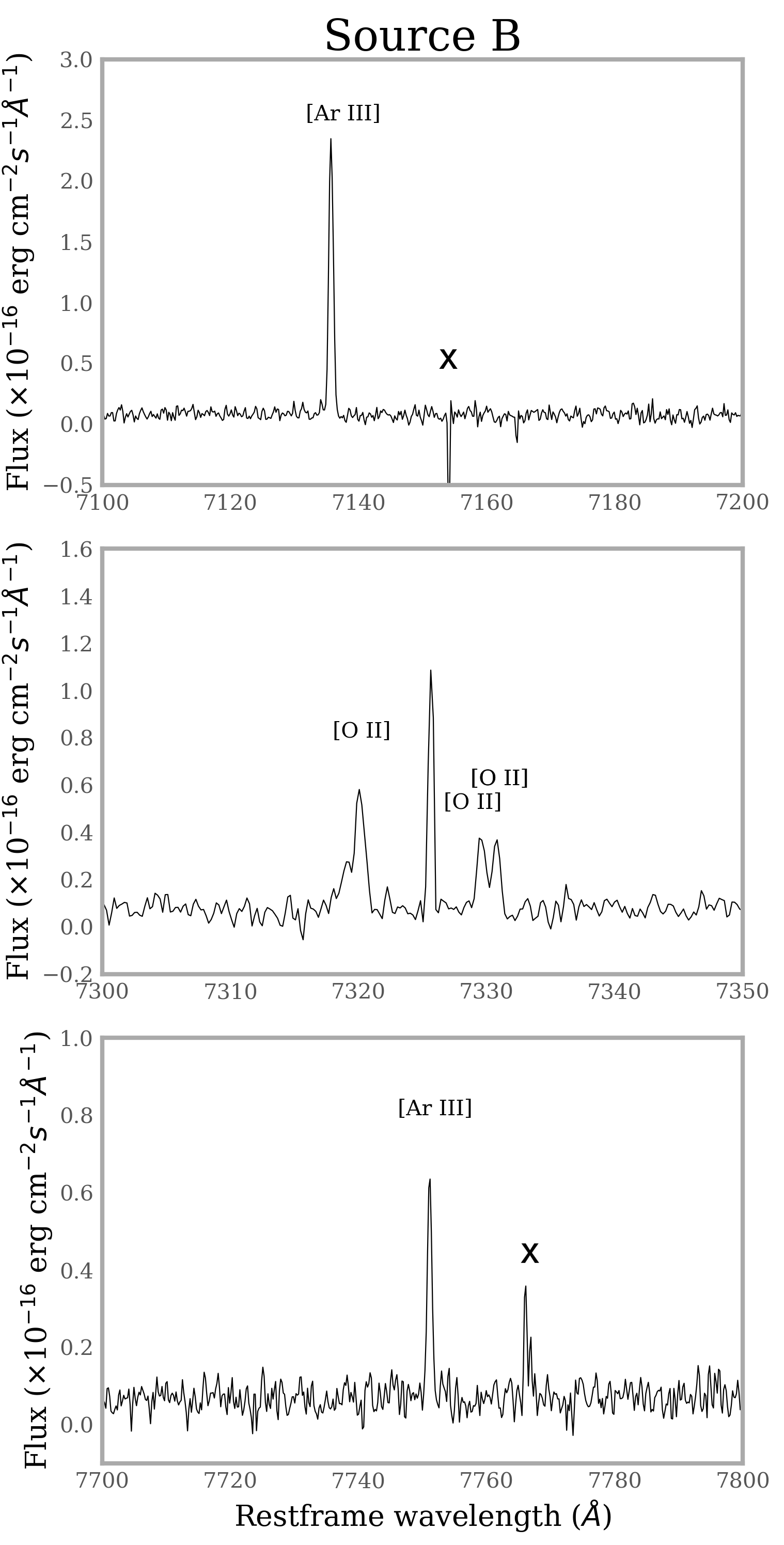}%
		\caption{Selected spectra showing strong emission lines from the VIS range of the X-Shooter spectra from Source B. The line classification is shown in the plot. The unlabeled line in the middle pannel is the unassociated 7325.71\AA\ line (see Table~\ref{tab:lines-wave}). Residuals of sky lines subtraction or spurious lines are marked with an x.}
        \label{fig:spectra-vis2}
\end{figure*}

\begin{figure*}
        \includegraphics[width=\textwidth]{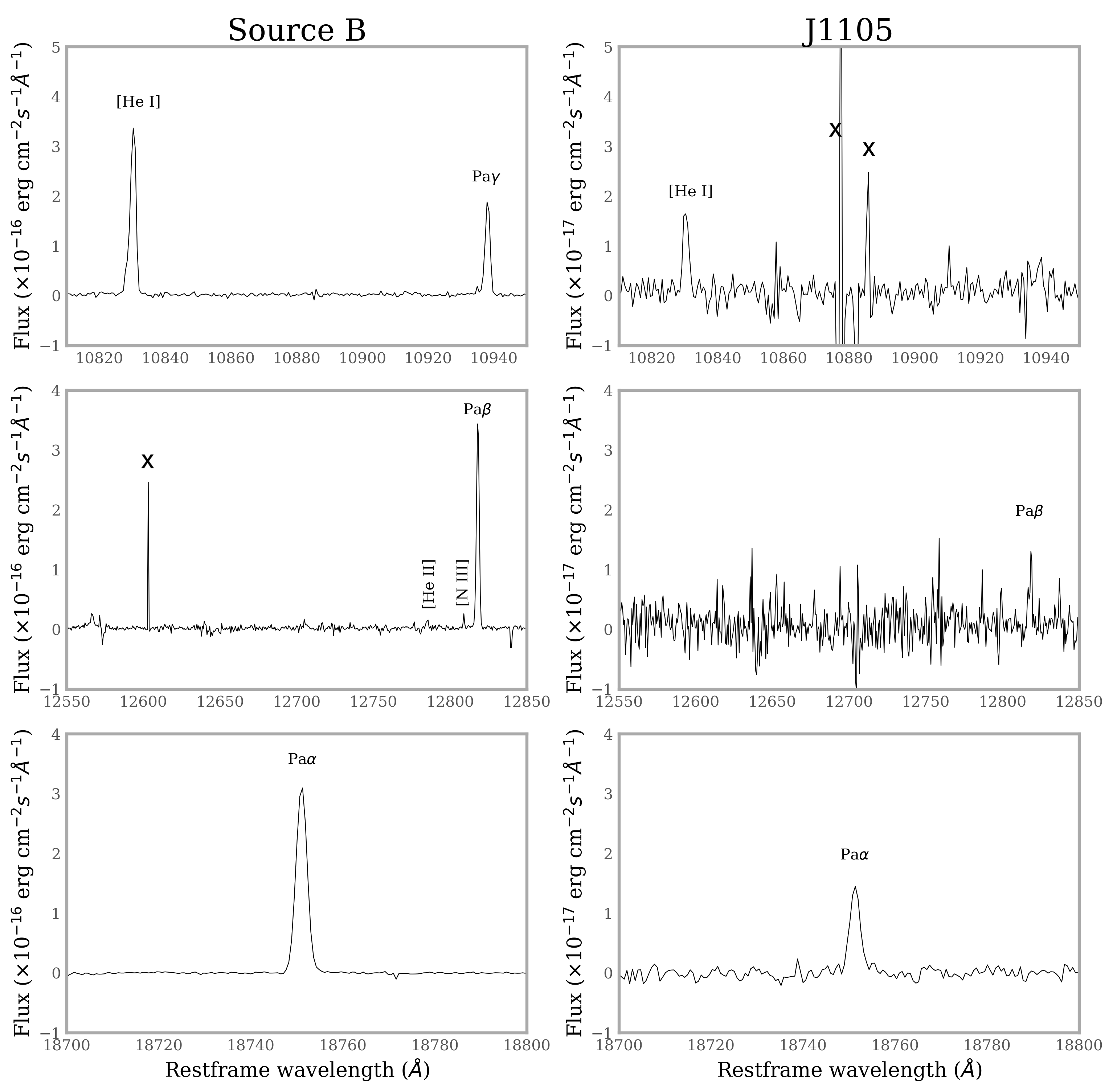}%
		\caption{Selected spectra showing strong emission lines from the NIR range of the X-Shooter spectra from Source B (left) and J1105 (right). The line classification is shown in the plot. Residuals of sky lines subtraction or spurious lines are marked with an x.}
        \label{fig:spectra-nir}
\end{figure*}

\begin{figure*}
        \includegraphics[width=\textwidth]{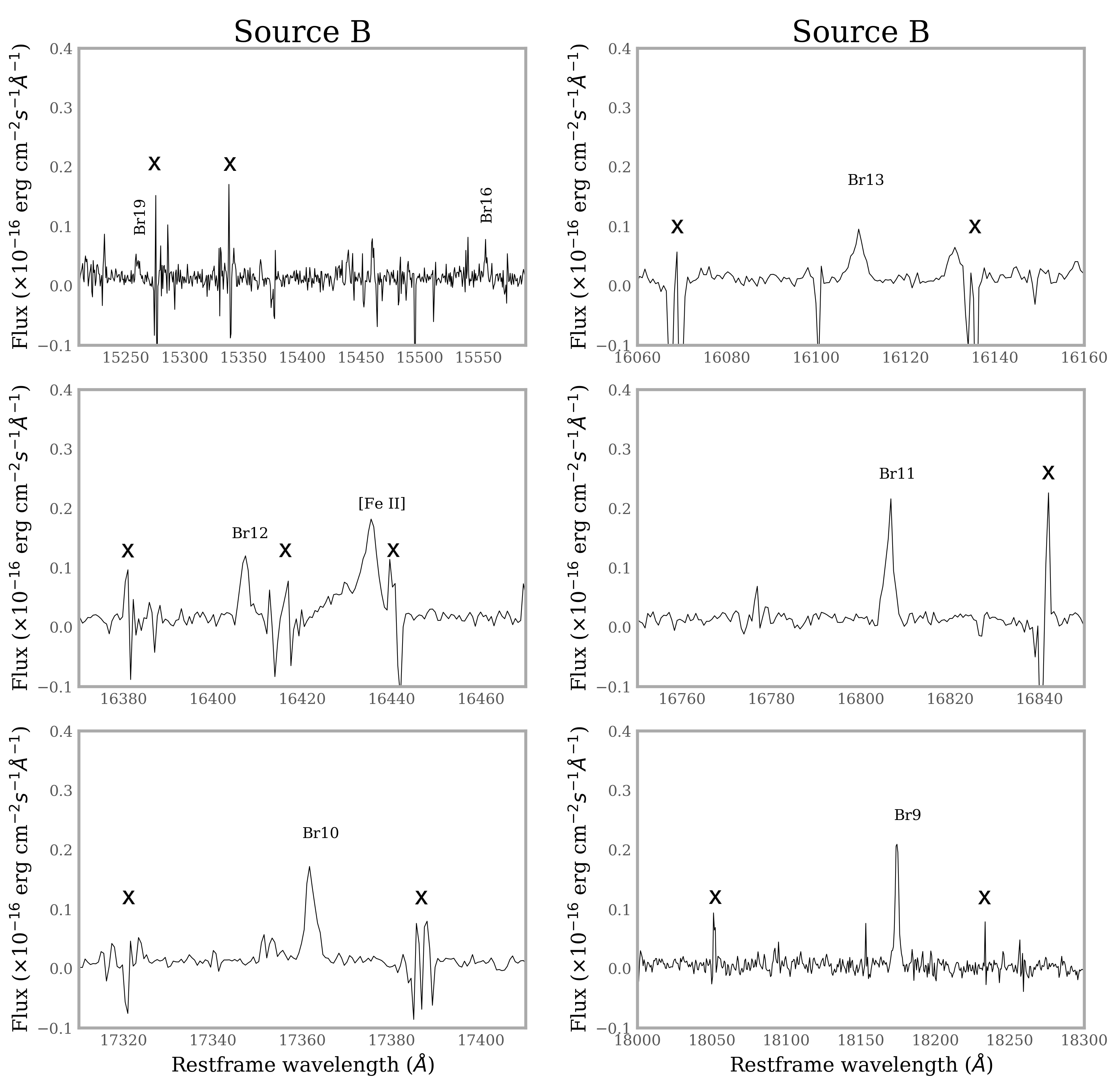}%
		\caption{Selected spectra showing strong emission lines from the NIR range of the X-Shooter spectra from Source B. The line classification is shown in the plot. Residuals of sky lines subtraction or spurious lines are marked with an x.}
        \label{fig:spectra-nir3}
\end{figure*}

\begin{figure*}
        \includegraphics[width=0.5\textwidth]{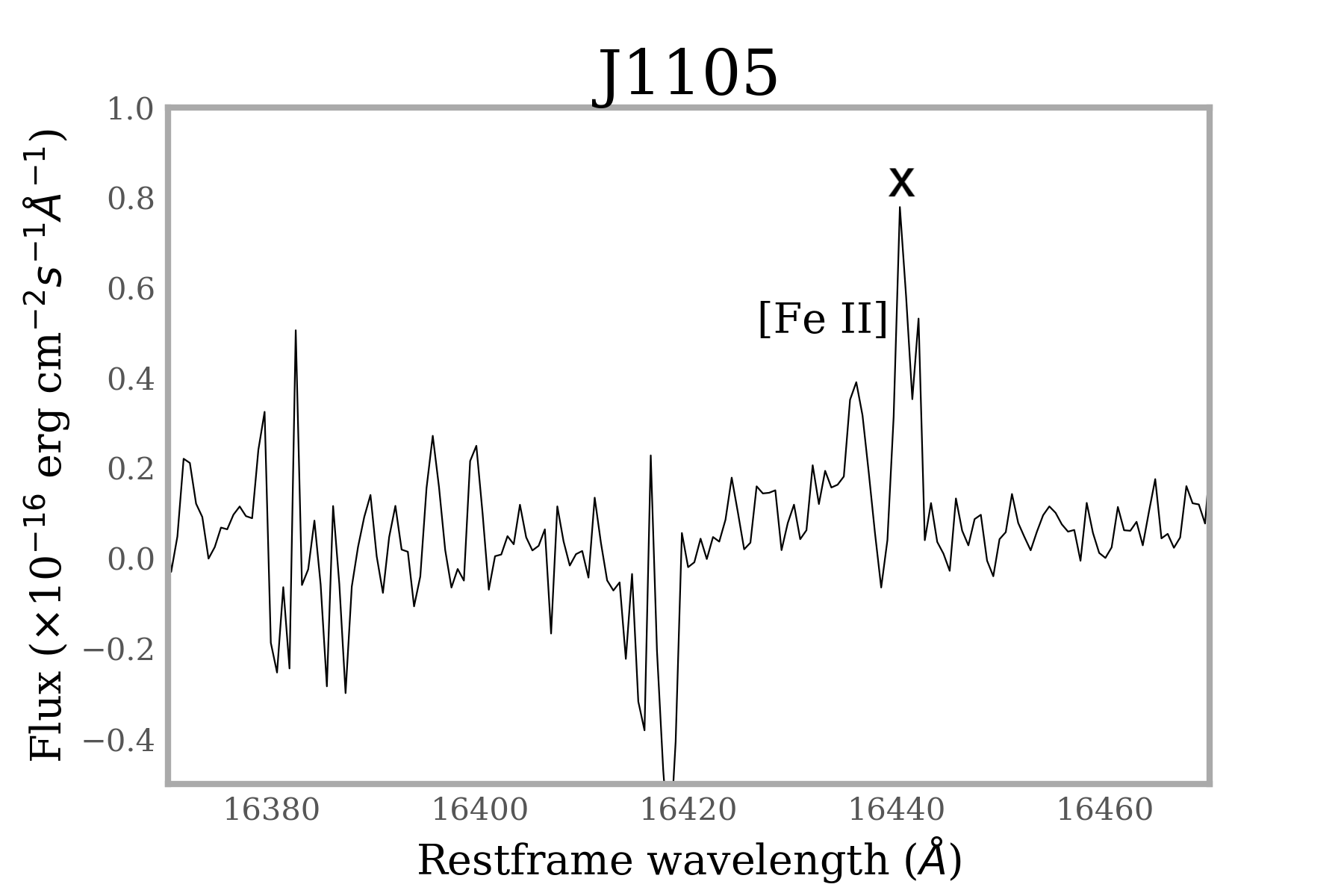}%
		\caption{Selected region from the NIR range of the X-Shooter spectra from J1105, showing the weak Fe {\scshape ii} line. Residuals of sky lines subtraction or spurious lines are marked with an x.}
        \label{fig:spectra-nir2}
\end{figure*}

%%%%%%%%%%%%%%%%%%%%%%%%%%%%%%%%%%%%%%%%%%%%%%%%%%

% Don't change these lines
\bsp	% typesetting comment
\label{lastpage}
\end{document}